\documentclass[12pt]{amsart}
\usepackage{amscd}     
\usepackage{amssymb}
\usepackage{amsmath, amsthm, graphics}
\usepackage{xypic}     
\LaTeXdiagrams        
\usepackage[all]{xy}
\xyoption{2cell} \UseAllTwocells \xyoption{frame} \CompileMatrices
\allowdisplaybreaks[3]
\usepackage{amsfonts}
\usepackage[normalem]{ulem}
\usepackage{hyperref}
\usepackage{tikz}
\usepackage{mathtools}
\usepackage{verbatim} 
\usepackage{upgreek}

\usepackage[none]{hyphenat}

\usepackage{latexsym}
\usepackage{epsfig}
\usepackage{amsfonts}
\usepackage{enumerate}
\usepackage{times}
\usepackage{bbm}

\newtheorem{theorem}{Theorem}[subsection]
\newtheorem{corollary}[theorem]{Corollary}
\newtheorem{lemma}[theorem]{Lemma}

\newtheorem{proposition}[theorem]{Proposition}

\newtheorem{definition}[theorem]{Definition}

\theoremstyle{remark}

\theoremstyle{remark}

\numberwithin{equation}{section}

\newcommand{\C}{\mathbb{C}}
\newcommand{\R}{\mathbb{R}}
\newcommand{\N}{\mathbb{N}}

\def\<{\left\langle}
\def\>{\right\rangle}

\newcommand{\ebar}{\overline{\varepsilon}}

\def\b1{{\mathbf 1}}

\begin{document}

\title[Finite Gap Conditions and Small Dispersion Asymptotics for Periodic Benjamin-Ono]{Finite Gap Conditions and Small Dispersion Asymptotics\\ for the Classical Periodic Benjamin-Ono Equation}

\author[A. Moll]{Alexander Moll}
\address{Department of Mathematics\\ 
Northeastern University \\ 
550 Nightingale\\ 
Boston, MA USA}
\email{a.moll@northeastern.edu}

\begin{abstract} In this paper we characterize the Nazarov-Sklyanin hierarchy for the classical periodic Benjamin-Ono equation in two complementary degenerations: for the multi-phase initial data (the periodic multi-solitons) at fixed dispersion and for bounded initial data in the limit of small dispersion.  First, we express this hierarchy in terms of a piecewise-linear function of an auxiliary real variable which we call a \textit{dispersive action profile} and whose regions of slope $\pm 1$ we call {gaps} and {bands}, respectively.  Our expression uses Kerov's theory of profiles and Kre\u{\i}n's spectral shift functions.  Next, for multi-phase initial data, we identify Baker-Akhiezer functions in Dobrokhotov-Krichever and Nazarov-Sklyanin and prove that multi-phase dispersive action profiles have finitely-many gaps determined by the singularities of their Dobrokhotov-Krichever spectral curves.  Finally, for bounded initial data independent of the coefficient of dispersion, we show that in the small dispersion limit, the dispersive action profile concentrates weakly on a convex profile which encodes the conserved quantities of the dispersionless equation.  To establish the weak limit, we reformulate Szeg\H{o}'s first theorem for Toeplitz operators using spectral shift functions.  To illustrate our results, we identify the dispersive action profile of sinusoidal initial data with a profile found by Nekrasov-Pestun-Shatashvili and its small dispersion limit with the convex profile found by Vershik-Kerov and Logan-Shepp.
\end{abstract}

  \maketitle

\setcounter{tocdepth}{1}
\tableofcontents

\pagebreak

\setcounter{section}{0}
\section{Introduction and Statement of Results} \label{SECIntro}

\noindent For $v = v(x,t)$, $v: \R \times [0, \infty) \rightarrow \R$, the classical Benjamin-Ono equation \cite{Benj, DavisAcrivos, Ono} is \begin{equation} \label{CBOE}  \partial_t v  + v \partial_x v = \tfrac{\overline{\varepsilon}}{2} J[ \partial^2_{x} v]  \end{equation} a non-linear, non-local integro-differential equation with dispersion coefficient $\ebar >0$ defined by the spatial Hilbert transform $(J\varphi) (x) = \text{P.V.} \tfrac{1}{\pi} \int_{- \infty}^{+\infty} \tfrac{\varphi(y) dy}{x-y}$ with $J e^{ \pm \textbf{i}kx} = \mp \textbf{i} e^{\pm \textbf{i}kx}$ and $J 1 = 0$.  Global well-posedness of (\ref{CBOE}) is established for rapidly-decaying $L^2$ initial data by Ionescu-Kenig \cite{IonescuKenig} and for periodic $L^2$ initial data by Molinet \cite{Molinet}.  We discuss the notation $\ebar$ in \textsection [\ref{SECSinusoidalDISCUSSION}] and write $v=v(x,t; \ebar)$ for solutions of (\ref{CBOE}).  For a comprehensive survey of research on (\ref{CBOE}) see Saut \cite{Saut2018}.\\
\\
\noindent In this paper we study an explicit family of infinitely-many conserved quantities for the classical Benjamin-Ono equation (\ref{CBOE}) with periodic initial data found by Nazarov-Sklyanin \cite{NaSk2}.  In \textsection [\ref{SECnazsklyINTRO}] we recall the classical Nazarov-Sklyanin hierarchy from \cite{NaSk2}.  In \textsection [\ref{SECdispersiveactionprofilesINTRO}] we present an alternative expression for the classical Nazarov-Sklyanin hierarchy through what we call \textit{dispersive action profiles} using Kerov's theory of profiles \cite{Ke1}.  In \textsection [\ref{SECFiniteGapResults}] we state our first result in Theorem [\ref{THEOREMMultiPhase}]: a characterization of dispersive action profiles for the $\ebar$-dependent multi-phase solutions (the periodic multi-solitons).  In \textsection [\ref{SECSmallDispersionLimitResults}] we state our second result in Theorem [\ref{THEOREMSmallDispersionLimit}]: a characterization of the small dispersion asymptotics of dispersive action profiles for arbitrary bounded $\ebar$-independent $v$.  We provide an outline of the paper in \textsection [\ref{SECoutline}] and give comments on previous results in \textsection [\ref{SECcomments}]. 

\subsection{Classical Nazarov-Sklyanin Hierarchy: All Baker-Akhiezer Averages are Conserved} \label{SECnazsklyINTRO}

\noindent We now define the classical Benjamin-Ono Lax operator and the Nazarov-Sklyanin hierarchy \cite{NaSk2}.
\begin{definition} \label{HardyDef} \textcolor{black}{For $w = e^{\textnormal{\textbf{i}} x}$ and $\mathbb{T} = \{w \in \C : |w|=1\}$,} the $L^2$-Hardy space $H_{\bullet}$ \textcolor{black}{on $\mathbb{T}$} is the \textcolor{black}{Hilbert space} closure of $\C[\textcolor{black}{w}]$ in $\textcolor{black}{H=}L^2(\mathbb{T})$.  Equivalently, with the {Szeg\H{o} projection}  \begin{equation} \label{HardyIntegralOperator} ( \uppi_{\bullet} \Phi)(w_+) =  \textcolor{black}{\textnormal{P.V.}} \textcolor{black}{\oint_{\mathbb{T}} \frac{ \Phi(w_-)  }{ w_- - w_+} \frac{ dw_- }{ 2 \pi \textnormal{\textbf{i}}}} , \end{equation} the periodic $L^2$-Hardy space  $H_{\bullet}$ is the image of $\uppi_{\bullet}$ applied to $H = L^2(\mathbb{T})$.
\end{definition}

\begin{definition} \label{DEFLaxOpIntro} For $\ebar >0$ and bounded real $v(x)$ $2\pi$-periodic in $x$ with Fourier coefficients $ V_k = \textcolor{black}{\int_0^{2 \pi}}e^{\textnormal{\textbf{i}} k x} v(x) \frac{ dx}{2\pi}$, the classical Lax operator $L_{\bullet}(v; \ebar)$ for the Benjamin-Ono equation \textnormal{(\ref{CBOE})} is the unique self-adjoint extension to Hardy space $H_{\bullet}$ of the essentially self-adjoint operator \begin{equation}\label{ClassicalLaxMatrix} L_{\bullet}(v; \ebar) \big |_{\C[\textcolor{black}{w}]} =
 \begin{bmatrix} 
(-0 \ebar + V_0 )& V_{-1} & V_{-2} & V_{-3} & \cdots &  \\ 
V_{1} & (-1 \ebar+ V_0)  & V_{-1} & V_{-2} &  \ddots &  \\
 V_{2} & V_{1} &  (-2 \ebar+ V_0)  & V_{-1} & \ddots & \\
 V_3 & V_2 & V_1 &(-3 \ebar + V_0 )& \ddots &   \\ 
  \vdots & \ddots & \ddots & \ddots & \ddots    \\ 
  \end{bmatrix}  \end{equation} \noindent presented in the basis $|h \rangle = \textcolor{black}{w^h}$ for $h=0,1,2,\ldots$ of $\C[\textcolor{black}{w}]$.  Equivalently, the Lax operator \begin{equation} \label{ClassicalLaxINTRINSIC} L_{\bullet}(v; \ebar)= - \ebar D_{\bullet} + L_{\bullet}(v) \end{equation} \noindent is the generalized Toeplitz operator of order $1$, where $L(v)$ is the operator of multiplication by $v$, $L_{\bullet}(v) = \uppi_{\bullet} L(v) \uppi_{\bullet}$ is the Toeplitz operator of symbol $v$, and $D_{\bullet}$ acts by $D_{\bullet}| h \rangle  = h | h \rangle$.  \end{definition}

\noindent For background on Toeplitz operators, see Deift-Its-Krasovsky \cite{DeiftItsKra}.  The classical Lax operator $L_{\bullet}(v; \ebar)$ is essentially self-adjoint on $\C[\textcolor{black}{w}]$ since it is a bounded perturbation of $D_{\bullet} = \uppi_{\bullet}( \textcolor{black}{w \partial_w} ) \uppi_{\bullet}$ by the Toeplitz operator $L_{\bullet}(v)$ of bounded symbol $v$.  The next two notions are from \cite{NaSk2}.

\begin{definition} \label{ClassicalNSbakerakhiezerDEF} For $u \in \C \setminus \R$ and $w \in \mathbb{T}$, the classical Baker-Akhiezer function \begin{equation} \label{ClassicalNSbakerakhiezer} \Phi^{BA}(u,w| v; \ebar) = \textcolor{black}{ \sum_{h=0}^{\infty} \langle h |} \frac{1}{u - L_{\bullet}(v; \ebar)} | 0 \rangle \textcolor{black}{w^h} \end{equation} \noindent is the image of $|0 \rangle = e^{\textnormal{\textbf{i}} 0 x}= 1$ in $H_{\bullet}$ under the resolvent of the classical Lax operator \textnormal{(\ref{ClassicalLaxINTRINSIC})}.  \end{definition}


\begin{definition} The classical Nazarov-Sklyanin hierarchy is the set of classical observables \begin{equation} \label{ClassicalNSgeneratingINTRO} \ \ \ \Phi_0^{BA}(u | v; \ebar)  = \oint_{\mathbb{T}} \Phi^{BA}(u,w | v; \ebar) \frac{dw}{ 2 \pi \textnormal{\textbf{i}} w} \end{equation} indexed by $u \in \C \setminus \R$ and defined as the circle averages of the Baker-Akhiezer function \textnormal{(\ref{ClassicalNSbakerakhiezer})}.
\end{definition}

\begin{theorem} \label{ClassicalNStheorem} \textnormal{[Nazarov-Sklyanin \cite{NaSk2}]} For any $\ebar >0$, any bounded real $v$, and any $u \in \C \setminus \R$, \begin{equation}\label{ClassicalNSResult} \frac{d}{dt} \Phi_0^{BA} (u | v(x,t;\ebar); \ebar)  = 0 \end{equation} \noindent the circle averages \textnormal{(\ref{ClassicalNSgeneratingINTRO})} are all conserved if $v$ evolves by the Benjamin-Ono equation \textnormal{(\ref{CBOE})}.\end{theorem}
\noindent \textcolor{black}{As we discuss in \textsection [\ref{SECdapDISCUSSION}], Theorem [\ref{ClassicalNStheorem}] was independently discovered by G\'{e}rard-Kappeler \cite{GerardKappeler2019}.}  For verification that the quantum Baker-Akhiezer function, quantum hierarchy, and Theorem 2 in Nazarov-Sklyanin \cite{NaSk2} reduce to the classical objects (\ref{ClassicalNSbakerakhiezer}), (\ref{ClassicalNSgeneratingINTRO}), and Theorem [\ref{ClassicalNStheorem}] presented here in the semi-classical limit $\hbar \rightarrow 0$, see \textsection 4 and \textsection 8 in \cite{Moll2}.

\subsection{Classical Nazarov-Sklyanin Hierarchy: Expression through Dispersive Action Profiles} \label{SECdispersiveactionprofilesINTRO} 

\noindent In this paper we characterize the conserved quantities (\ref{ClassicalNSgeneratingINTRO}) in the small dispersion limit and for the known multi-phase solutions of (\ref{CBOE}).  To do so, we first establish the following result for (\ref{ClassicalNSgeneratingINTRO}).

\begin{proposition} \label{ClassicalNSreexpression} For $\ebar >0$ and bounded real $v$, there is some ${n_{\star}}(v) \in \{0,1,2,3,\ldots\} \cup \{\infty\}$ so that the classical Nazarov-Sklyanin hierarchy \textnormal{(\ref{ClassicalNSgeneratingINTRO})} of Baker-Akhiezer averages \begin{equation} \label{SimpleMeromorphicFormula} \Phi_0^{BA} ( u | v; \ebar) = \prod_{i=1}^{{n_{\star}}(v)} \frac{ u - S_i^{\downarrow}( v; \ebar)}{ u - S_{i-1}^{\uparrow} ( v; \ebar)}  \end{equation}
is meromorphic \textcolor{black}{in $u \in \C$} with simple \textcolor{black}{interlacing real} zeroes $\{S_i^{\downarrow}(v; \ebar)\}_{i=1}^{\infty}$ and poles $\{S_{i}^{\uparrow} (v; \ebar) \}_{i=0}^{\infty}$ \begin{equation} \label{InterlacingINTROyay} \cdots < S_i^{\uparrow} (v; \ebar) < S_i^{\downarrow}( v; \ebar) < \cdots < S_1^{\uparrow}(v; \ebar) < S_1^{\downarrow}(v; \ebar) < S_0^{\uparrow}(v; \ebar) \end{equation} \noindent with no accumulation point except $-\infty$ and so that 
\begin{equation} \label{AverageINTROyay} \lim_{n \rightarrow {n_{\star}}(v)} \Big ( S_n^{\uparrow} (v; \ebar) + \sum_{i=1}^n |S_i^{\downarrow}(v; \ebar) - S_{i-1}^{\uparrow}(v; \ebar) | \Big ) = \frac{1}{2\pi}  \int_0^{2\pi} v(x) dx . \end{equation} \end{proposition}
\noindent We prove Proposition [\ref{ClassicalNSreexpression}] in \textsection [\ref{SECNazarovSklyanin}] as an application of general results from Kerov's theory of profiles \cite{Ke1} and Kre\u{\i}n's theory of spectral shift functions which we review in \textsection [\ref{SECJacobiKerov}].  To make contact with Kerov's theory of profiles, we use (\ref{SimpleMeromorphicFormula}) to express (\ref{ClassicalNSgeneratingINTRO}) in terms of a profile $f( c| v; \ebar)$.
\begin{definition} \label{DispersiveActionProfileDEF} For $\ebar >0$ and bounded real $v$, the dispersive action profile $f( c | v; \ebar)$ is the piecewise-linear function of $c \in \R$ with $|f'(c)|=1$ almost everywhere, strictly interlacing local minima $S_i^{\uparrow}(v; \ebar)$ and local maxima $S_i^{\downarrow} (v; \ebar)$ in \textnormal{(\ref{InterlacingINTROyay})}, and $f(c | v; \ebar) \sim |c- a|$ as $c \rightarrow \pm \infty$ for $a = V_0 = \int_0^{2\pi} v(x) \frac{dx}{2\pi}$.  For such $f(c | v; \ebar)$, the classical Nazarov-Sklyanin hierarchy \textnormal{(\ref{ClassicalNSgeneratingINTRO})} is
\begin{equation} \label{ClassicalNSviaDAP} \Phi_0^{BA} ( u | v; \ebar) = \textnormal{exp} \Bigg ( \int_{- \infty}^{+\infty} \textnormal{log} \Bigg [ \frac{1}{u-c} \Bigg ] \tfrac{1}{2} f''(c | v; \ebar) dc \Bigg ) .\end{equation}

\end{definition}
\begin{figure}[htb]
\centering
\includegraphics[width=0.5 \textwidth]{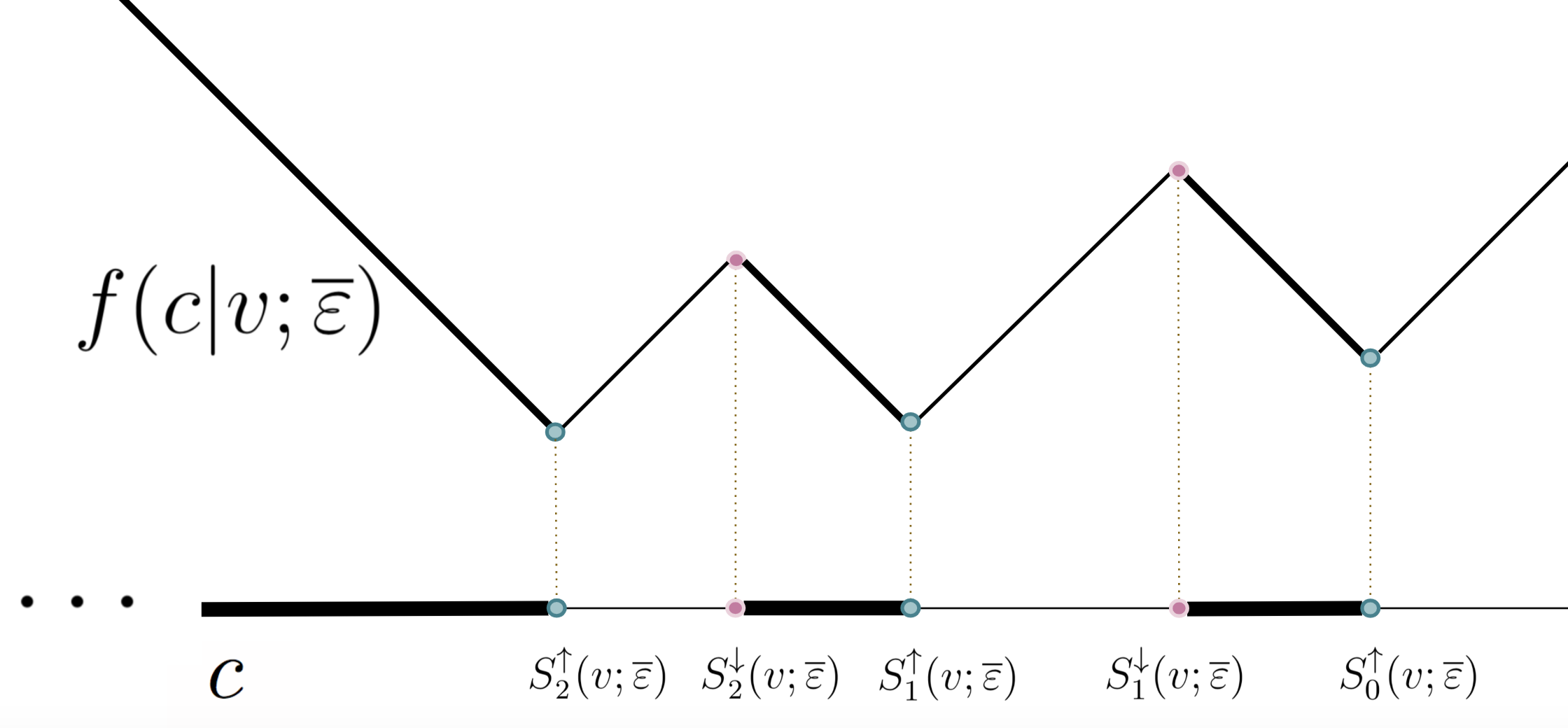}
\caption{Dispersive action profile $f(c| v; \ebar)$ of generic smooth periodic real $v$ with typically infinitely-many non-empty gaps $(S_i^{\uparrow} (v; \ebar), S_i^{\downarrow}(v; \ebar))$ extending to $- \infty$.}
\label{DispersiveActionProfileFIGURE}
\end{figure}

\begin{definition} \label{BandDef} The bands of dispersive action profiles are $[S_i^{\downarrow}(v; \ebar), S_{i-1}^{\uparrow}(v; \ebar)]$ where $f' = -1$. \end{definition}
\begin{definition} \label{GapDef} The gaps of dispersive action profiles are $(S_i^{\uparrow}(v; \ebar), S_i^{\downarrow} (v; \ebar))$ where $f' =+1$. \end{definition}

\noindent In \cite{Moll2} we identify gaps in dispersive action profiles with actions in a symplectic phase space of (\ref{CBOE}).  For details, see \textsection [\ref{SECdapDISCUSSION}].  Note ${n_{\star}}(v)$ in Proposition [\ref{ClassicalNSreexpression}] is the number \textcolor{black}{of non-empty gaps}.

\subsection{Statement of First Result: Dispersive Action Profiles of Multi-Phase Data are Finite Gap} \label{SECFiniteGapResults}

\noindent Satsuma-Ishimori \cite{SatsumaIshimori1979} discovered that the classical Benjamin-Ono equation (\ref{CBOE}) has multi-phase solutions, the periodic analogs of the multi-soliton solutions of (\ref{CBOE}) found by Matsuno \cite{Matsuno1979}.  Dobrokhotov-Krichever \cite{DobrokhotovKrichever} gave the following formula for the multi-phase solutions from \cite{SatsumaIshimori1979}.
\begin{definition} \label{DEFmultiphase} For $n=0,1,2\ldots$ with $2n+1$ real parameters $\vec{s} \in \R^{2n+1}$ ordered as \begin{equation} \label{BOMultiPhaseParameters} s_n^{\uparrow} < s_n^{\downarrow} < \cdots <  s_1^{\uparrow} < s_1^{\downarrow} < s_0^{\uparrow} \end{equation} \noindent and $n$ phases $\chi_n, \ldots, \chi_1 \in \mathbb{R}$ denoted $\vec{\chi} \in \mathbb{R}^n$, the multi-phase ($n$-phase) solutions of \textnormal{(\ref{CBOE})} are \begin{equation} \label{BOmPhaseSolution} v^{\vec{s}, \vec{\chi}} (x, t; \ebar) = s_n^{\uparrow} - \sum_{i=1}^n (s_{i-1}^{\uparrow} - s_i^{\downarrow}) - 2 \textcolor{black}{\ebar} \textnormal{Im}  \partial_x \log \det  M^{\vec{s}, \vec{\chi}} (x,t; \ebar )\end{equation} \noindent where $M^{\vec{s}, \vec{\chi}}(x,t;\ebar)$ is the $n \times n$ matrix with entries $M^{\vec{s}, \vec{\chi}}_{ij} (x,t; \ebar)$ for $1 \leq i,j \leq n$ defined by \begin{equation} \label{BOmPhaseSolutionMatrix} M^{\vec{s}, \vec{\chi}}_{ij} (x,t; \ebar)=\tfrac{1}{ s_{i-1}^{\uparrow} - s_j^{\downarrow}} \Big ( - 1 +  \delta(i-j) Z_i(\vec{s}) e^{  { \textnormal{\textbf{i}} ( \tfrac{s_{i-1}^{\uparrow} - s_i^{\downarrow} }{\ebar}) ( x - \chi_i - \tfrac{1}{2}(s_i^{\downarrow} + s_{i-1}^{\uparrow})t ) } } \Big ) \end{equation}  \begin{equation} Z_i(\vec{s}) =  \sqrt{ \tfrac{ s_{i-1}^{\uparrow} - s_n^{\uparrow} }{ s_{i}^{\downarrow} - s_n^{\uparrow}} } \prod_{j \neq i} \sqrt{ \tfrac{ (s_i^{\downarrow} - s_j^{\downarrow})( s_{i-1}^{\uparrow} - s_{j-1}^{\uparrow})}{ (s_{i-1}^{\uparrow} - s_j^{\downarrow}) ( s_i^{\downarrow} - s_{j-1}^{\uparrow})}} .\end{equation} 
\end{definition}

\noindent For $n=1$, (\ref{BOmPhaseSolution}) is the 1-phase periodic traveling wave found by Benjamin \cite{Benj} and Ono \cite{Ono}: \begin{equation} \label{BO1PhaseForm} v^{\vec{s}, \vec{\chi}}(x,t ; \ebar) = \frac{ (s_0^{\uparrow} - s_1^{\downarrow})^2}{ (s_1^{\downarrow} - s_1^{\uparrow}) + (s_0^{\uparrow} - s_1^{\uparrow}) - 2 \sqrt{ \frac{ s_0^{\uparrow} - s_1^{\uparrow}}{ s_1^{\downarrow} - s_1^{\uparrow}}} \cos \Big ( \big ( \frac{ s_0^{\uparrow} - s_1^{\downarrow}}{\ebar} \big ) \big (x - \chi_1 - \tfrac{1}{2}(s_1^{\downarrow} + s_0^{\uparrow}) t \big ) \Big )} .\end{equation}  \noindent In Dobrokhotov-Krichever \textnormal{\cite{DobrokhotovKrichever}}, parameters \textnormal{(\ref{BOMultiPhaseParameters})} \textcolor{black}{appear as} singularities of rational spectral curves.  We now encode these singularities $\vec{s}$ in a profile $f( c | \vec{s})$ a priori unrelated to the dispersive action profile $f( c | v; \ebar)$ and state our first result: the classical multi-phase solutions are finite gap.
\begin{definition} \label{MultiPhaseProfileDEF} For any $\vec{s} \in \R^{2n+1}$ satisfying \textnormal{(\ref{BOMultiPhaseParameters})}, the \textcolor{black}{Dobrokhotov-Krichever} profile $f(c |\vec{s})$ is the piecewise-linear function of $c \in \R$ with $|f'(c)|=1$ almost everywhere, interlacing local minima $s_i^{\uparrow}$ and local maxima $s_i^{\downarrow}$, and $f(c | \vec{s}) \sim |c- a|$ as $c \rightarrow \pm \infty$ for $a = s_n^{\uparrow} + \sum_{i=1}^n | s_i^{\downarrow} - s_{i-1}^{\uparrow} |$.\end{definition}

\begin{theorem} \label{THEOREMMultiPhase} For $\ebar > 0$, $\vec{s}$ in \textnormal{(\ref{BOMultiPhaseParameters})}, and $\vec{\chi} \in \R^n$ so $v^{\vec{s}, \vec{\chi}}(x,t; \ebar)$ in \textnormal{(\ref{BOmPhaseSolution})} is $2\pi$-periodic in $x$, \begin{equation} \label{FiniteGapWhatWeNeed} f(c | v^{\vec{s}, \vec{\chi}} (x, t; \ebar) ; \ebar) = f(c | \vec{s}) \end{equation} \noindent the multi-phase dispersive action profile coincides with the Dobrokhotov-Krichever profile and thus have finitely-many non-empty gaps.  Equivalently, the quantities in \textnormal{Proposition  [\ref{ClassicalNSreexpression}]} specialize to \begin{eqnarray}\label{finitegapIndex} {n_{\star}}(v^{\vec{s}, \vec{\chi}} (x,t; \ebar) ) &=& n \\ S_i^{\uparrow}(v^{\vec{s}, \vec{\chi}}(x, t; \ebar) ; \ebar) &=& s_i^{\uparrow} \\ S_i^{\downarrow}(v^{\vec{s}, \vec{\chi}}(x, t; \ebar) ; \ebar) &=& s_i^{\downarrow}. \end{eqnarray}
\end{theorem}

\noindent We prove Theorem [\ref{THEOREMMultiPhase}] in \textsection [\ref{SECfirstresultFINITEGAP}] by identifying the classical Baker-Akhiezer function (\ref{ClassicalNSbakerakhiezer}) of Nazarov-Sklyanin \cite{NaSk2} with the classical Baker-Akhiezer function in Dobrokhotov-Krichever \cite{DobrokhotovKrichever}.  As a companion to our Theorem [\ref{THEOREMMultiPhase}], we also prove in \textsection [\ref{SECfirstresultFINITEGAP}] that the dispersive action profiles $f(c | - v^{\vec{s}, \vec{\chi}} (x,t; \ebar); \ebar)$ of reflected multi-phase solutions $v \mapsto -v$ are not finite gap: \begin{proposition} \label{THEOREMLaurentFiniteGapPARTIALCONVERSE} For any $\ebar >0$, $\vec{s}$ in \textnormal{(\ref{BOMultiPhaseParameters})}, $\vec{\chi} \in \R^n$, unlike \textnormal{(\ref{finitegapIndex})}, ${n_{\star}}(-v^{\vec{s}, \vec{\chi}}(x,t;\ebar)) = \infty$. \end{proposition}

\noindent \textcolor{black}{In \textsection [\ref{SUBSECfinitegap}], we discuss the agreement of these results with subsequent work of G\'{e}rard-Kappeler \cite{GerardKappeler2019}.}

\subsection{Statement of Second Result: Dispersive Action Profiles at Small Dispersion are Convex} \label{SECSmallDispersionLimitResults} 

\noindent Our second result is that the small dispersion limit $\ebar \rightarrow 0$ of conserved quantities (\ref{ClassicalNSgeneratingINTRO}) for (\ref{CBOE}) of Nazarov-Sklyanin \cite{NaSk2} are conserved for the classical {dispersionless Benjamin-Ono equation} \begin{equation} \label{CRHE} \partial_t v + v \partial_x v = 0 \end{equation} at $\ebar = 0$.  While solutions $v(x,t;0)$ to (\ref{CRHE}) with differentiable periodic initial data do not remain continuous for all time $t$, for small $t$ the method of characteristics implies that for any $c \in \R$ \begin{equation} \label{RayleighFunctionConvexActionProfile} F(c|v;0) = \int_0^{2 \pi} {1}_{\{v(x,t;0) \leq c\}}(x) \tfrac{dx}{2\pi} \end{equation}
\noindent the measure of $\{x \in [0,2\pi] : v(x,t;0) \leq c\}$ is conserved.  We now state this fact using profiles.  \begin{definition} \label{DEFConvexActionProfile} For any bounded $v(x)$ $2\pi$-periodic in $x$, the {convex action profile} $f(c| v;0)$ is the convex function of $c \in \R$ characterized by $F(c| v;0) = \tfrac{1+f'(c|v;0)}{2}$ for $F(c|v;0)$ in \textnormal{(\ref{RayleighFunctionConvexActionProfile})} and $f(c | v;0) \sim |c- a|$ as $c \rightarrow \pm \infty$ where $a = \int_0^{2\pi} v(x) \frac{dx}{2 \pi}$.\end{definition}  \begin{proposition} \label{PROPConvexActionProfileConserved} For $c \in \R$, differentiable $2\pi$-periodic initial data $v(x,0;0)$, and small $t$, the $c$-value of the convex action profile is conserved $\partial_t f(c | v(x,t;0);0) = 0 $ if $v(x,t;0)$ solves \textnormal{(\ref{CRHE})}. \end{proposition} \pagebreak

 \noindent We review Proposition [\ref{PROPConvexActionProfileConserved}] in \textsection[\ref{SECwhereIreview}].  In terms of the convex action profile $f(c| v; 0)$, we show:
\begin{theorem} \label{THEOREMSmallDispersionLimit} For $\ebar>0$ and any $v(x)$ real, $2\pi$-periodic in $x$, bounded, and independent of $\ebar$, the dispersive action profile $f(c | v; \ebar)$ for \textnormal{(\ref{ClassicalNSgeneratingINTRO})} converges weakly to the convex action profile $f( c | v; 0)$ in the small dispersion limit in the following sense: for any $u \in \C \setminus \R$ fixed, as $\ebar \rightarrow 0$ we have \begin{equation} \label{SmallDispersionWeakLimit} \int_{-\infty}^{+\infty} \log \Bigg [ \frac{1}{u-c} \Bigg ] \tfrac{1}{2} f''(c | v; \ebar) dc \rightarrow \ \int_{-\infty}^{+\infty} \log \Bigg [ \frac{1}{u-c} \Bigg ] \tfrac{1}{2} f''(c | v; 0)dc.  \end{equation}
\end{theorem}
 \begin{figure}[htb]
\centering
\includegraphics[width=0.45 \textwidth]{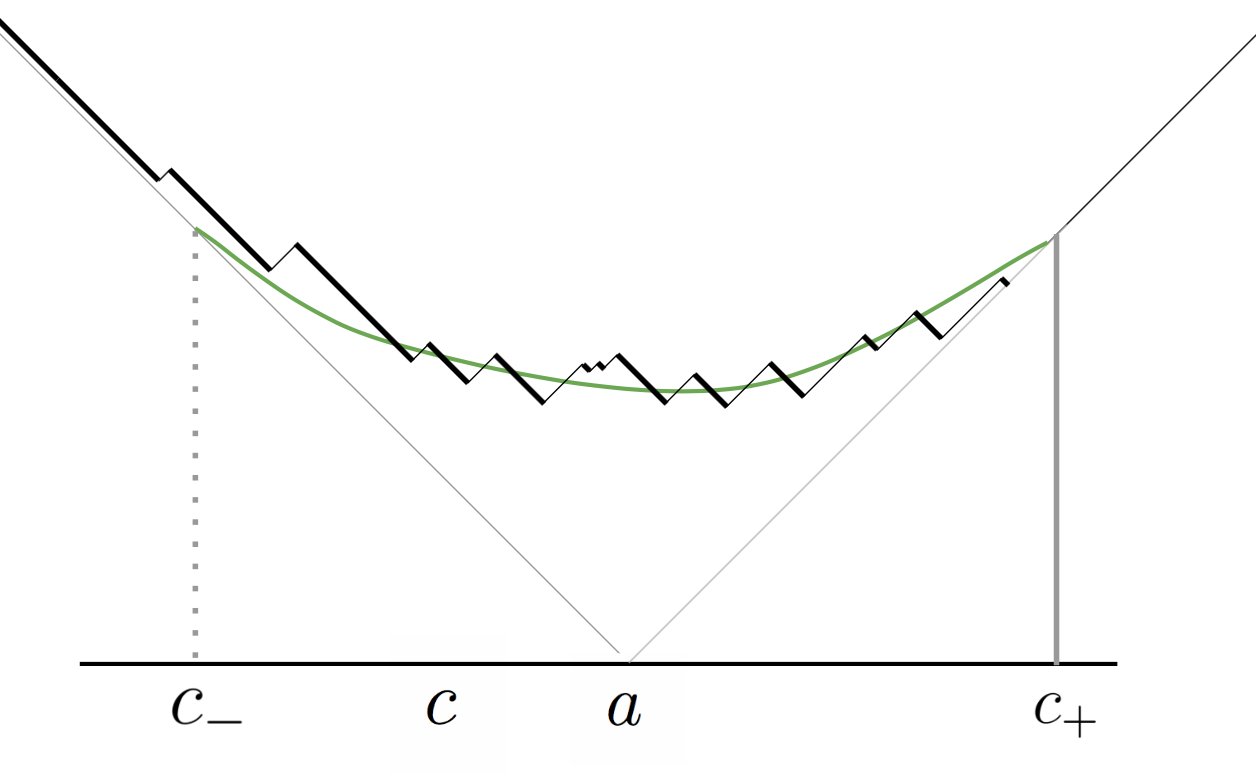}
\caption{Dispersive action profile and convex action profile for generic $v$ in Theorem [\ref{THEOREMSmallDispersionLimit}] with $a= \int_0^{2\pi} v(x) \tfrac{dx}{2 \pi}$, $c_- = \inf_x v(x)$, and $c_+ = \sup_x v(x)$.}
\label{ProfileBOSmallDispersionLimitTEMP}
\end{figure}

\noindent We prove Theorem [\ref{THEOREMSmallDispersionLimit}] in \textsection[\ref{SECSmallDispersionLimit}].  \textcolor{black}{Theorem [\ref{THEOREMSmallDispersionLimit}] gives small dispersion asymptotics of dispersive action profiles, not of solutions to (\ref{CBOE}).} The key ingredient in our proof is a non-local interpretation of the local conservation laws (\ref{RayleighFunctionConvexActionProfile}) from the method of characteristics which we achieve by reformulating Szeg\H{o}'s First Theorem for Toeplitz operators using spectral shift functions in \textsection [\ref{SECToeplitz}].  Note: the assumptions in Theorem [\ref{THEOREMSmallDispersionLimit}] exclude the $\ebar$-dependent multi-phase solutions (\ref{BOmPhaseSolution}), which by Theorem [\ref{THEOREMMultiPhase}] have dispersive action profiles $f( c | \vec{s})$ independent of $\ebar$ and non-convex.


\subsection{Outline} \label{SECoutline}  In \textsection [\ref{SECJacobiKerov}] we review Kerov's theory of profiles \cite{Ke1} and Kre\u{\i}n's theory of spectral shift functions \cite{KreinSSF} \textcolor{black}{so as to prove Proposition [\ref{KMKstatementII}], a relationship between profiles and spectral shift functions of a certain class of unbounded self-adjoint operators which extends a result of Kerov \cite{Ke1}.}  Using Proposition [\ref{KMKstatementII}], in \textsection [\ref{SECNazarovSklyanin}] we prove Proposition [\ref{ClassicalNSreexpression}] and derive our expression (\ref{ClassicalNSviaDAP}) for the classical Nazarov-Sklyanin hierarchy (\ref{ClassicalNSgeneratingINTRO}) in terms of dispersive action profiles.  In \textsection\ref{SECfirstresultFINITEGAP} we prove Theorem [\ref{THEOREMMultiPhase}] -- that the classical multi-phase solutions (\ref{BOmPhaseSolution}) of (\ref{CBOE}) have dispersive action profiles with finitely-many non-empty gaps -- and also Proposition [\ref{THEOREMLaurentFiniteGapPARTIALCONVERSE}] -- that after reflection $v \mapsto -v$, multi-phase solutions are no longer finite gap.  In \textsection [\ref{SECToeplitz}] we give a non-local characterization of the convex action profile $f(c|v;0)$ for bounded $v$ from Definition [\ref{DEFConvexActionProfile}] using Szeg\H{o}'s First Theorem for Toeplitz operators.  In \textsection [\ref{SECSmallDispersionLimit}], we prove Theorem [\ref{THEOREMSmallDispersionLimit}], that for bounded real $v$, in the small dispersion limit the dispersive action profiles converge to the convex action profiles.  In \textsection [\ref{SECSinusoidal}], to illustrate our results, we identify the dispersive action profile $f( c | v_{\star}; \ebar)$ for sinusoidal initial data \begin{equation} \label{SinusoidalInitialData} v_{\star}(x) = 2 \cos x \end{equation} \noindent with a profile in Nekrasov-Pestun-Shatashvili \cite{NekPesSha} and identify its convex small dispersion limit $f( c | v_{\star}; 0)$ with the profile in Vershik-Kerov \cite{KeVe} and Logan-Shepp \cite{LoSh}.  \textcolor{black}{In \textsection [\ref{SECmotivation}], we discuss our results in the larger context of the Whitham modulation theory for classical dispersive shock waves solutions to (\ref{CBOE}).  In \textsection [\ref{SECcomments}] we give comments on our results and compare them to previous results.}


\section{Kre\u{\i}n Spectral Shift Functions in Kerov's Theory of Profiles} \label{SECJacobiKerov}

\noindent In this section, we review Kerov's theory of profiles \cite{Ke1} in order to prove in Proposition [\ref{KMKstatementII}] that certain Kre\u{\i}n spectral shift functions \cite{KreinSSF} define profiles, extending a result of Kerov \cite{Ke1}.
\subsection{Profiles: Interlacing Measures and Shifted Rayleigh Functions} \label{SECProfilesSRF} \noindent We follow Kerov \cite{Ke1}.

\begin{definition} \label{ProfileDefinition} A \textit{{profile}} is a function $f : \R \rightarrow \R$ of $c \in \R$ which is $1$-Lipshitz \begin{eqnarray} \label{profileDefLipshitz} |f (c_1) - f(c_2) | & \leq & | c_1 - c_2 |  \ \ \   \ \ \ \ \ \ \ \  \text{for all} \ \ \ \ c_1, c_2 \in \R \end{eqnarray}

\noindent and whose derivatives $f'(c) \rightarrow \pm 1$ as $c \rightarrow \pm \infty$ so that \begin{equation} \label{profileDefDecay} \int_{- \infty}^{0} (1 + f'(c))\cdot  \frac{ dc}{1+|c|} < \infty \ \ \ \ \  \text{and}  \ \ \ \ \  \int_{0}^{+\infty} (1- f'(c) ) \cdot \frac{ dc}{ 1+|c|} < \infty . \end{equation} \noindent Let ${\mathbf{P}^{\vee}}$ denote the space of all profiles.
\end{definition} 

 \begin{definition} The {Rayleigh function} $F_f: \R \rightarrow [0,1]$ of a profile $f $ is defined by \begin{equation} F_f(c) := \tfrac{1}{2} (1 + f'(c) ) .\end{equation} \end{definition}

\begin{definition} Rayleigh functions $F_f$ of bounded variation define {Rayleigh measures} $dF_f$. \end{definition}

\begin{definition} Non-negative measures $dF^{\uparrow}, dF^{\downarrow}$ on $\R$ are {interlacing measures} if their difference $dF^{\uparrow} - dF^{\downarrow} = dF_f$ is a Rayleigh measure of some profile $f$.  In this case, write $dF^{\uparrow}_f, dF^{\downarrow}_f$.
\end{definition}

\begin{definition} A profile $f$ is of {compact support} if $dF_f$ exists and has compact support. \end{definition}

  \begin{figure}[htb]
\centering
\includegraphics[width=0.5 \textwidth]{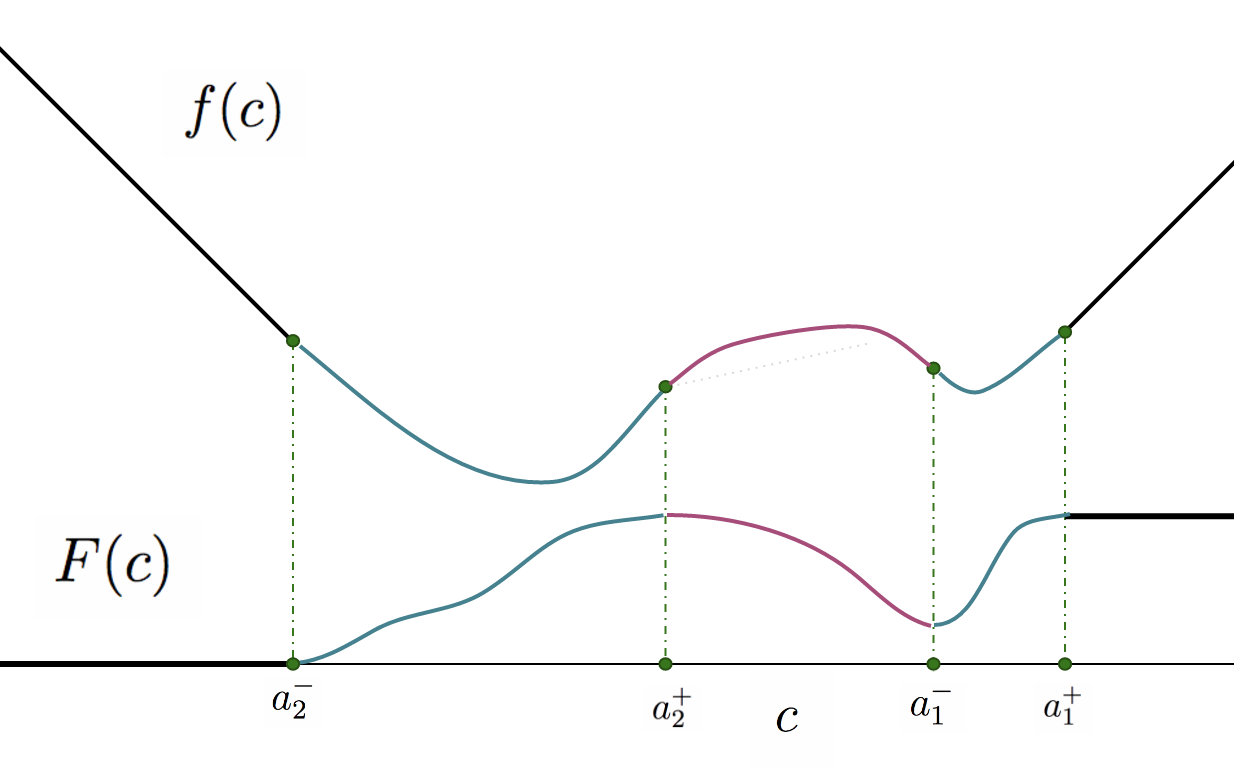}
\caption{A profile $f$ of compact support $[a_2^-, a_1^+]$ and its Rayleigh function $F_f$ with interlacing measures $dF^{\uparrow}_f$, $dF_f^{\downarrow}$ supported on $[a_2^-, a_2^+] \cup [a_1^-, a_1^+]$ and $[a_2^+, a_1^-]$.}
\label{FigureGenericProfile}
\end{figure}
\noindent The points of inflection of $f$ separate regions of convexity and concavity, which correspond to increasing $\uparrow$ or decreasing $\downarrow$ regions of the Rayleigh function $F_f$, hence to our notation for interlacing measures.  For all profiles, we consider their behavior relative to the profile $f_0(c) = |c- 0|$:
\begin{definition} \label{DEFShiftedRayleighFunction}The {shifted Rayleigh function} $\xi_f$ of a profile $f$ is the difference \begin{equation} \label{SRFformula} \ \ \ \ \ \ \ \  \xi_f(c) := F_f(c) - F_{f_0}(c) \end{equation} between its Rayleigh function $F_f(c)$ and the Rayleigh function $F_{f_0}(c) = \mathbbm{1}_{[0, \infty)}(c)$ of $f_0(c) = |c|$. \end{definition} \noindent Note that one can recover $f$ from $\xi_f$ by $f(c) = \int_{- \infty}^c \xi_f(y) dy + \int_c^{+\infty} (1 - \xi_f(y)) dy$.

\subsection{Profiles: Convex Profiles and Profiles of Interlacing Sequences}

\noindent Kerov's profiles interpolate between \textit{convex profiles} - such as the convex action profile $f(c |v;0)$ in Definition [\ref{DEFConvexActionProfile}] - and the \textit{profiles of interlacing sequences} - such as the dispersive action profile $f( c|v; \ebar)$ in Definition [\ref{DispersiveActionProfileDEF}].
\begin{definition} \label{ConvexProfileDEFINITION} A {convex profile} is a profile $f$ of bounded variation with $dF_f^{\downarrow}(c)=0$, i.e. whose Rayleigh measure $dF_f(c) = dF^{\uparrow}_f(c)$ is an arbitrary probability measure $dF_f$.
\end{definition}   
  \begin{figure}[htb]
\centering
\includegraphics[width=0.40 \textwidth]{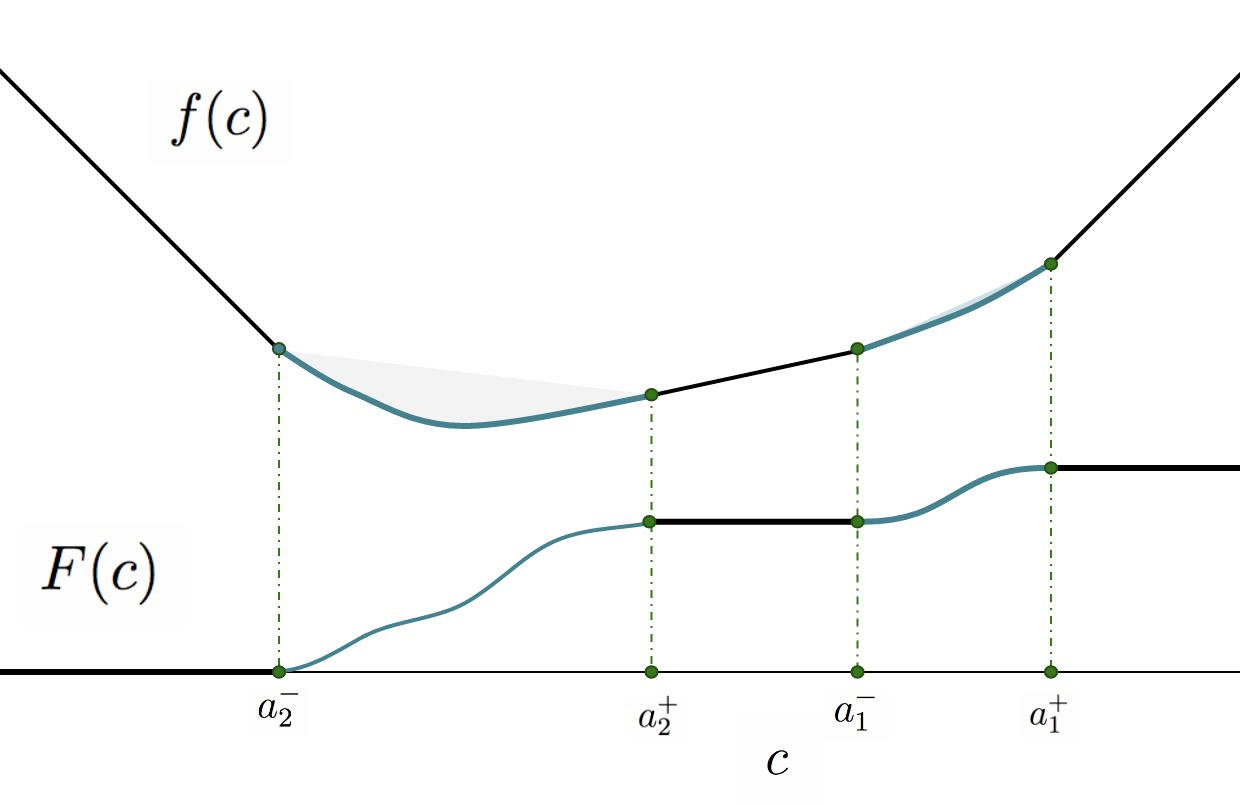}
\caption{Convex profile with $dF_f = dF_f^{\uparrow}$ supported on $[a_2^-, a_2^+]$ and $[a_1^-, a_1^+]$.}
\label{FigureConvexProfile}

\end{figure}

 \begin{definition} \label{ProfileInterlacingSequencesDEFINITION} A {profile of interlacing sequences} is any profile with Rayleigh measure \begin{equation} dF_f(c) = \sum_{i=0}^{n} \delta(c - s_i^{\uparrow}) - \sum_{i=1}^{n} \delta( c - s_i^{\downarrow}). \end{equation} \noindent where $n_{\star} \in \N \cup \{\infty\}$ and $\{s_i^{\uparrow} \}_{i=0}^{n_{\star}}$, $\{s_i^{\downarrow} \}_{i=1}^{n_{\star}}$ strictly interlace \begin{equation} \label{InterlacingSequenceFormula} \cdots < s_i^{\uparrow} < s_{i}^{\downarrow} <  \cdots < s_{1}^{\uparrow} < s_1^{\downarrow} < s_0^{\uparrow}. \end{equation} \end{definition}

 \noindent Dobrokhotov-Krichever profiles of Definition [\ref{MultiPhaseProfileDEF}] are profiles of interlacing sequences (\ref{BOMultiPhaseParameters}). 

  \begin{figure}[htb]
\centering
\includegraphics[width=0.40 \textwidth]{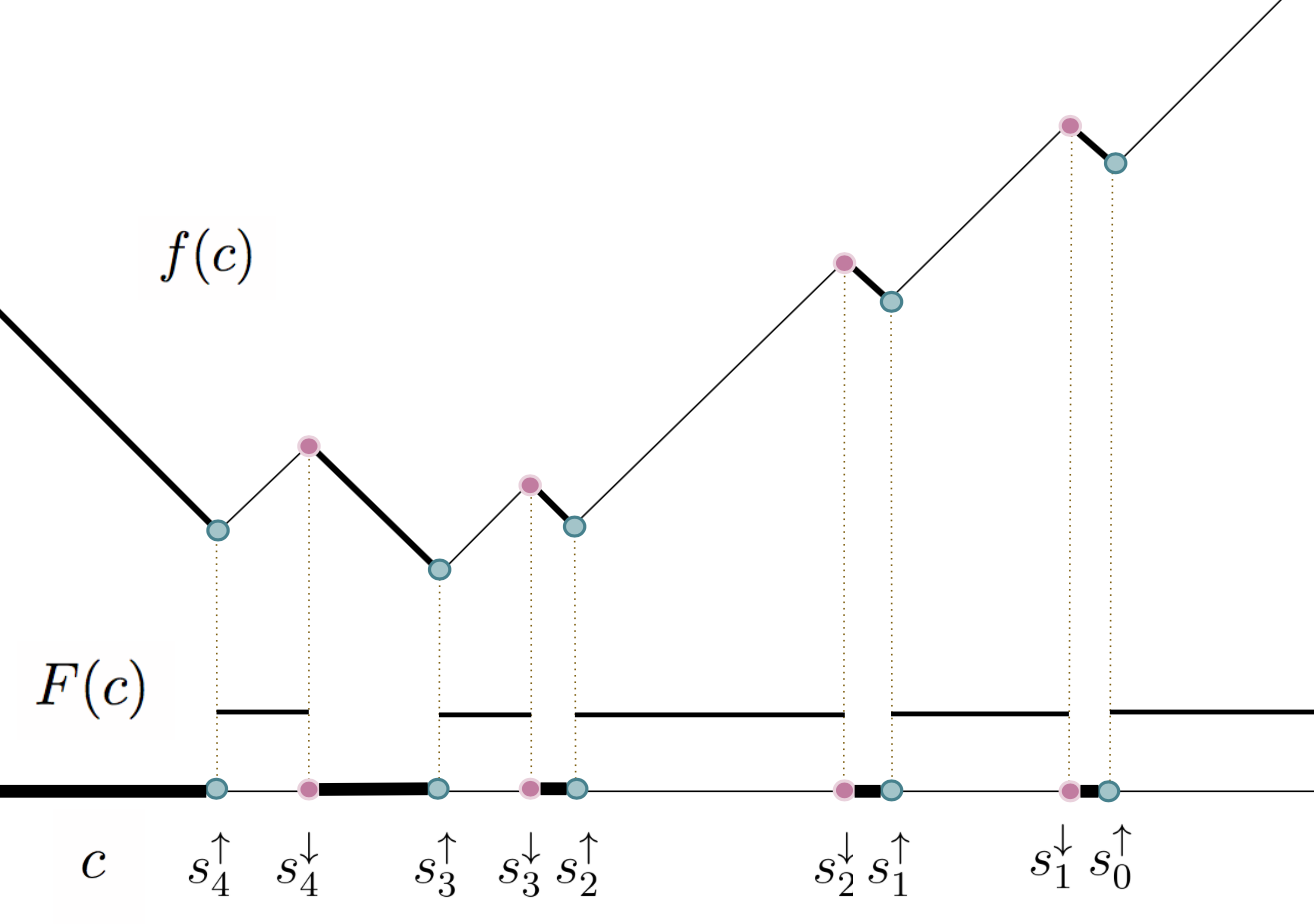}
\caption{A profile $f$ of interlacing sequences with $n=4$ bands of finite length.}
\label{FigureInterlacingSequenceProfile}
\end{figure}
\noindent Whereas regions of concavity and convexity of a generic profile may be of full Lebesgue measure, for profiles of interlacing sequences the regions of convexity and concavity are localized at the local minima and maxima of the piecewise-linear profile $f$.  Most importantly, every $a \in \R$ defines $f_a( c) = |c- a|$ that is both convex and also the profile of the interlacing sequences $\{a\}, \emptyset$ with $2n+1$ interlacing local extrema for $n=0$.  In fact, $f_a(c) = f(c | a; 0) = f(c | a; \ebar)$ is both the convex and dispersive action profile for $v \equiv a$ the constant $n=0$ phase solution of (\ref{CBOE}) with $\vec{s} = s_0^{\uparrow} = a$.


\pagebreak

\subsection{Profiles: Transition Measures and Kerov's Markov-Kre\u{\i}n Correspondence} \label{SECKMK} \noindent The decay conditions in Definition [\ref{ProfileDefinition}] of profiles were chosen carefully by Kerov in \cite{Ke1} to formulate his Markov-Kre\u{\i}n correspondence, a bijection between profiles $f$ and probability measures $\tau$ on $\R$.
\begin{definition} For $u \in \C \setminus \R$, the {$T^{\uparrow}$-observable} of a profile $f \in \mathbf{P}^{\vee}$ is defined in terms of its shifted Rayleigh function $\xi_f$ from \textnormal{Definition [\ref{DEFShiftedRayleighFunction}]} for $u \in \C \setminus \R$ by \begin{equation} \label{TUpObservableDefinition}
 T^{\uparrow}(u) \Big |_{f} = \frac{1}{u} \cdot  \textnormal{exp} \Bigg ( - \int_{- \infty}^{+\infty} \frac{ \xi_f(c) dc}{u-c} \Bigg ). \end{equation}
\end{definition}

\begin{proposition} If a profile $f \in \mathbf{P}^{\vee}$ is of bounded variation, integration by parts implies that its $T^{\uparrow}$-observable can be written through its Rayleigh measure $dF_f = \tfrac{1}{2} f''(c) dc$ for $u \in \C \setminus \R$ by \begin{equation} T^{\uparrow}(u) \Big |_{f} = \textnormal{exp} \Bigg ( \int_{- \infty}^{+\infty} \log \Bigg [ \frac{1}{u-c} \Bigg ] \tfrac{1}{2} f''(c) dc \Bigg ). \end{equation}
\end{proposition}

\begin{definition} Let $\mathbf{P}$ denote the space of probability measures on $\R$. \end{definition}

\begin{theorem} \label{KMKstatement} \textnormal{(Kerov's Markov-Kre\u{\i}n Correspondence \cite{Ke1})}  The $T^{\uparrow}$-observable $T^{\uparrow}(u)|_f$ defined for any profile $f \in \mathbf{P}^{\vee}$ and any $u \in \C \setminus \R$ by \textnormal{(\ref{TUpObservableDefinition})} is also the Stieltjes transform  \begin{equation}  \label{KMKgeneral} \int_{- \infty}^{+\infty} \frac{ d \tau_f^{\uparrow}(c) }{ u-c} =  T^{\uparrow}(u)  |_f = \frac{1}{u} \cdot  \textnormal{exp} \Bigg ( - \int_{- \infty}^{+\infty} \frac{ \xi_f(c) dc}{u-c} \Bigg )  \end{equation}

\noindent of a unique probability measure $d\tau^{\uparrow}_f \in \mathbf{P}$ called the {transition measure} of $f$.  Moreover, the formula \textnormal{(\ref{KMKgeneral})} defining $f \mapsto d\tau_f^{\uparrow}$ is a bijection $\mathbf{P}^{\vee} \rightarrow {\mathbf{P}}$ between profiles and probability measures on $\R$.\end{theorem} \noindent Our notation $\uparrow$ in $T^{\uparrow}(u)|_f$ emphasizes a relationship between the transition measure $d\tau^{\uparrow}_f$ and the $dF_f^{\uparrow}$ in the Jordan decomposition of the Rayleigh measure $dF_f$.  For example, if $f = f_{\vec{s}}$ is the profile of an interlacing sequence $s_{n}^{\uparrow} <  s_{n}^{\downarrow} < \cdots < s_1^{\uparrow} < s_1^{\downarrow} < s_0^{\uparrow}$ for some $n$, equation (\ref{KMKgeneral}) becomes \begin{equation} \label{TakePartialFracOfThis} \int_{- \infty}^{+\infty} \frac{ d \tau_{f_{\vec{s}}}^{\uparrow}(c)}{u-c} = T^{\uparrow}(u) |_{f_{\vec{s}}} = \frac{ \prod_{i=1}^{n} (u - s_i^{\downarrow}) }{ \prod_{i=0}^n (u - s_i^{\uparrow})} .\end{equation}

\noindent The partial fraction decomposition of (\ref{TakePartialFracOfThis}) shows that the transition measure $d\tau^{\uparrow}_f$ is non-negative and supported on $\{s_n^{\uparrow}, \ldots, s_1^{\uparrow}, s_0^{\uparrow}\}$ the support of $dF^{\uparrow}(c)$.  For general profiles, Theorem [\ref{KMKstatement}] implies an important relationship between the moments and supports of $dF_f$ and $d \tau^{\uparrow}_f$. 
\begin{corollary} \label{KMKsupport} \textnormal{[Kerov \cite{Ke1} \textsection 2.3]} In the Markov-Kre\u{\i}n correspondence \textnormal{(\ref{KMKgeneral})}, the support of the Jordan component $dF^{\uparrow}_f$ of the Rayleigh measure $dF_f$ and the transition measure $d\tau_f^{\uparrow}$ coincide, and the $T^{\uparrow}$-observable has a $1/u$ expansion \begin{equation}\label{KMKpolynomialrelation} \sum_{\ell=0}^{\infty} (T_{\ell}^{\uparrow})\big |_f u^{- \ell -1} = T^{\uparrow}_f(u) = \textnormal{exp} \Bigg (\textcolor{black}{ \sum_{p = 1}^{\infty} (O_p)\big |_f \frac{ u^{-p}}{p}} \Bigg ) \end{equation}

\noindent where $(T_{\ell}^{\uparrow})|_f = \int_{- \infty}^{+\infty} c^{\ell} d \tau^{\uparrow}$ and \textcolor{black}{$(O_p)|_f= - p \int_{- \infty}^{+\infty} c^{p-1} \xi_f(c ) dc$, so $O_p$ is a universal polynomial independent of $f$ in $T^{\uparrow}_{\ell}$ for $0 \leq \ell \leq p$. Moreover, $(O_p)|_f = \int_{- \infty}^{+\infty} c^p d \xi_f( c)$} if $f$ is bounded variation.\end{corollary}

\pagebreak

\subsection{Jacobi Operators: Embedded Principal Minors and Titchmarsh-Weyl Functions} \label{SECJacobiOperatorsTWFunctions} 
\noindent We now gather basic notions of Jacobi operators whose spectral theory we discuss in the next subsection.
\begin{definition} \label{PerturbationDeterminantDEF}Given a pair $L_{\bullet},L_{+}$ of possibly unbounded self-adjoint operators on a Hilbert space $H_{\bullet}$ with $L_{\bullet} - L_+$ trace class, the {perturbation determinant} of $u-L_+$ with respect to $u-L_{\bullet}$ \begin{equation} \label{DefFredDef} \frac{ \det_{H_{\bullet}} (u - L_+) }{ \det_{H_{\bullet}} (u -L_{\bullet}) } :=  \det\nolimits_{H_{\bullet}} \big ( \mathbbm{1} + (L_{\bullet}- L_+)( u - L_{\bullet})^{-1} \big ) \end{equation}
\noindent is well-defined by the Fredholm determinant for any $u \in \C \setminus \R$. \end{definition} \noindent The left side of (\ref{DefFredDef}) is a well-defined ratio of Fredholm determinants in the case that both $L_+$ and $L_{\bullet}$ are trace class, and under these assumptions can be shown to be equal to the right side of (\ref{DefFredDef}).  For the much more general case of arbitrary self-adjoint $L_{\bullet}$, $L_+$ with $L_{\bullet} - L_+$ trace class, we regard the left side of (\ref{DefFredDef}) as notation for the right side of (\ref{DefFredDef}) which is well-defined for such $L_{\bullet}, L_+$.\\
\\
\noindent We now turn to a particular rank $2$ perturbation $L_+$ of a generic $L_{\bullet} $. \begin{definition} \label{MinorGeneralDefinitions} For $\psi_0 \in H_{\bullet}$, let $H_{0} = \C | \psi_0 \rangle$ and consider the decomposition $H_{\bullet} = H_{0} \oplus H_{+}$ with orthogonal projections $\uppi_{0}: H_{\bullet} \rightarrow H_{0}$, $\uppi_{+}: H_{\bullet} \rightarrow H_{+}$.  The {principal $(\psi_0, \psi_0)$-{minor} of $L_{\bullet}$} \begin{equation}  L_+^{\perp} =\uppi_{+} L_{\bullet}  \uppi_{+}  \end{equation}
\noindent acts in $H_{+}$ while the {embedded principal $(\psi_0, \psi_0)$-{minor}} $L_{+}$ acts in $H_{\bullet} \cong H_{0} \oplus H_{+}$ by \begin{equation} L_+  = 0 \oplus L_+^{\perp}. \end{equation} \end{definition}

\noindent The distinction between the principal minor $L_+^{\perp}$ and the embedded principal minor $L_+$ is crucial for us in this paper, as the perturbation determinant (\ref{DefFredDef}) is only well-defined for a pair of operators defined on the same space $H_{\bullet}$ such as $L_+$ and $L_{\bullet}$ but not $L_+^{\perp}$ and $L_{\bullet}$.  The next result is well-known in the spectral theory of orthogonal polynomials on the real line \cite{SimonSzego}. 
\begin{theorem} \label{OPRL} If $\psi_0 \in H_{\bullet}$ is {cyclic} for $L_{\bullet}$ and $L_{\bullet} |_{\psi_0}$ the restriction of $L_{\bullet}$ to its dense orbit is {essentially self-adjoint}, then the $(\psi_0, \psi_0)$-matrix element of the resolvent is the $\tfrac{1}{u}$ multiple of
\begin{equation} \label{OPRLformula} \langle \psi_0 | \frac{1}{u - L_{\bullet}}| \psi_0 \rangle  = T^{\uparrow}(u) \big |_{L_{\bullet};\psi_0}  = \frac{1}{u} \cdot \frac{ \det_{H_{\bullet}} (u -L_+) }{ \det_{H_{\bullet}} ( u - L_{\bullet}) } \end{equation}

\noindent the {perturbation determinant} of the embedded principal minor $u-L_+$ with respect to $u-L_{\bullet}$ in $H_{\bullet}$. \end{theorem}

\noindent Theorem [\ref{OPRL}] follows from truncating $L_{\bullet}$, writing the result as a tri-diagonal Jacobi matrix, and using Cramer's rule.  To apply Theorem [\ref{OPRL}] in practice, one must check that the restriction of the operator $L_{\bullet}$ to the $L_{\bullet}$-orbit of $\psi_0$ is essentially self-adjoint.  A large class of such $L_{\bullet}$ are the \textit{bounded} self-adjoint operators.  As one sees in the proof of Theorem [\ref{OPRL}], the Galerkin approximation to $L_{\bullet}$ is a Jacobi matrix in a particular basis, so $L_{\bullet}$ may be viewed as a \textit{one-sided Jacobi operator}. \begin{definition} \label{TWDefinition} The {Titchmarsh-Weyl function} of a Jacobi operator $L_{\bullet}$ with cyclic $\psi_0$ is the function $T^{\uparrow}(u)|_{L_{\bullet}, \psi_0}$ of $u \in \C \setminus \R$ defined by either side of formula \textnormal{(\ref{OPRLformula})}. \end{definition} \noindent $T^{\uparrow}(u)$ is also known as the {Titchmarsh-Weyl m-function} in the theory of Jacobi operators \cite{SimonSzego}.  
\pagebreak

\subsection{Jacobi Operators: Spectral Measures and Spectral Shift Functions} \label{SECSpectralMeasuresSSF}
\noindent We now convert the equality (\ref{OPRLformula}) of matrix elements of Jacobi operators $L_{\bullet}, L_+$ into a statement in spectral theory.
\begin{definition} \label{SpectralMeasureDef} The {spectral measure} of $L_{\bullet}$ at $\psi_0$ is the probability measure $d\tau^{\uparrow} \in \mathbf{P}$ defined by \begin{equation} \label{SpectralMeasureFORMULA} \int_{- \infty}^{+\infty} \frac{ d \tau^{\uparrow} (c)}{ u-c} = \langle \psi_0 | \frac{1}{u - L_{\bullet}} | \psi_0 \rangle \end{equation}
\noindent for every $u \in \C \setminus \R$.  To emphasize its definition, we write $d \tau^{\uparrow}(c) = d \tau^{\uparrow}_{\psi_0, \psi_0}(c | L_{\bullet})$.
\end{definition} \noindent For trace-class perturbations, there is a relative notion of spectral measure due to Kre\u{\i}n \cite{KreinSSF}.
\begin{definition} \label{SSFDEFohyeah} Given any pair $L_{\bullet}, L_+$ of possibly unbounded self-adjoint operators on a Hilbert space $H_{\bullet}$ so that $L_{\bullet} - L_+$ is trace class, the {spectral shift function} $\xi(c | L_{\bullet}, L_+)$ is defined for all $u \in \C \setminus \R$ by the perturbation determinant in \textnormal{Definition [\ref{PerturbationDeterminantDEF}]} according to the formula \begin{equation} \label{SSFformula} \frac{ \det_{H_{\bullet}} ( u - L_+) }{ \det_{H_{\bullet}} (u - L_{\bullet}) } = \textnormal{exp} \Bigg ( - \int_{- \infty}^{+\infty} \frac{\xi(c | L_{\bullet}, L_+) dc}{u-c } \Bigg ) . \end{equation}
\end{definition}
\begin{theorem} \label{LifshitzKreinTraceFormula} \textnormal{[Lifshitz-Krein Trace Formula]} If $\phi: \R \rightarrow \C$ has $\phi'(c)$ with Fourier transform in $L^1(\R)$, for $L_{\bullet}, L_+$ possibly unbounded self-adjoint operators with $L_{\bullet} - L_+$ trace class, one has \begin{equation} \label{LKtraceformula} \textnormal{Tr}_{H_{\bullet}}  \Big [ \phi(L_{\bullet}) - \phi(L_+) \Big ] = - \int_{- \infty}^{+\infty} \phi'(c) \xi(c | L_{\bullet}, L_+) dc \end{equation} \noindent which simplifies to $\int_{- \infty}^{+\infty} \phi(c) d \xi(c | L_{\bullet}, L_+)$ if $\xi$ has bounded variation. \end{theorem}
\noindent For review of spectral shift functions and (\ref{LKtraceformula}), see Birman-Pushnitski \cite{BirPush} and Birman-Yafaev \cite{BirYaf}.
\begin{corollary} \label{OPRLcorollary}Under the assumptions of essential self-adjointness in Theorem \textnormal{[\ref{OPRL}]}, the spectral measure $d\tau^{\uparrow}_{\psi_0, \psi_0}(c | L_{\bullet})$ of $L_{\bullet}$ at $\psi_0$ determines the spectral shift function $\xi(c | L_{\bullet}, L_+)$ of $L_{\bullet}, L_+$ by \begin{equation} \label{OPRLcorollaryFormula} \int_{- \infty}^{+\infty} \frac{d \tau^{\uparrow}(c)}{u-c} = \frac{1}{u} \cdot \textnormal{exp} \Bigg ( - \int_{- \infty}^{+\infty} \frac{\xi(c | L_{\bullet}, L_+) dc}{u-c } \Bigg ) .\end{equation} \end{corollary}

\noindent Formula (\ref{OPRLcorollaryFormula}) is a particular case of formula (\ref{KMKgeneral}).  By Corollary [\ref{OPRLcorollary}], we have proven:

\begin{proposition} \label{KMKstatementII} Under the assumptions of essential self-adjointness of $L_{\bullet} |_{\psi_0}$ in Theorem \textnormal{[\ref{OPRL}]}, a self-adjoint operator $L_{\bullet}$ and  $\psi_0 \in H_{\bullet}$ determine a unique profile $f \in \textbf{P}^{\vee}$ so that \begin{itemize}
\item The $T^{\uparrow}$-observable $T^{\uparrow}(u) |_f$ in \textnormal{(\ref{TUpObservableDefinition})} is the Titchmarsh-Weyl function \textnormal{(\ref{OPRLformula})}
\item The transition measure $\tau_f^{\uparrow}$ in \textnormal{(\ref{KMKgeneral})} is the {spectral measure} of $L_{\bullet}$ at $\psi_0$ in \textnormal{(\ref{SpectralMeasureFORMULA})}
\item The shifted Rayleigh function $\xi_f$ in \textnormal{(\ref{SRFformula})} is the \textcolor{black}{\textit{spectral shift function}} $\xi (c | L_{\bullet}, L_+)$ in \textnormal{(\ref{SSFformula})}. \end{itemize}
\end{proposition}

\noindent Our Proposition [\ref{KMKstatementII}] generalizes the case of $L_{\bullet}$ bounded proved by Kerov in \textsection5-\textsection6 of \cite{Ke1}.  Our distinction between the Rayleigh function $F_f$ and shifted Rayleigh function $\xi_f = F - \mathbbm{1}_{[0, \infty)}$ corrects the statement of Theorem 6.1.3 in \cite{Ke1} by accounting for the difference between the principal minor $L_+^{\perp}$ on $H_{+} = H_{\psi_0}^{\perp}$ and the embedded principal minor $L_+ = 0 \oplus L_+^{\perp}$ on $H_{\bullet}$. Finally, note that the converse of our Proposition [\ref{KMKstatementII}] is not true: the only probability measures $d \tau^{\uparrow}$ arising as spectral measures of such essentially self-adjoint $L_{\bullet}|_{\psi_0}$ at $\psi_0$ are those whose Hamburger moment problem is determinate, a result of Nevanlinna discussed by Simon in \cite{SimonCMP}.  Hamburger indeterminate $d \tau^{\uparrow}$ still determine a unique profile $f$ by Theorem [\ref{KMKstatement}], just not in the manner of Proposition [\ref{KMKstatementII}].

\section{Lax Spectral Shift Functions and Dispersive Action Profiles} \label{SECNazarovSklyanin} 

\noindent In this section we prove Proposition [\ref{ClassicalNSreexpression}] and derive our expression (\ref{ClassicalNSviaDAP}) for the classical Nazarov-Sklyanin hierarchy (\ref{ClassicalNSgeneratingINTRO}) in terms of dispersive action profiles $f( c | v; \ebar)$ in Definition [\ref{DispersiveActionProfileDEF}].

\subsection{Classical Benjamin-Ono Lax Operator: Principal Minor and Embedded Principal Minor}

\noindent To apply results from \textsection [\ref{SECJacobiOperatorsTWFunctions}] in \textsection [\ref{DAPproof}], we first specialize the definition of principal minors $L_{+}^{\perp}$ and embedded principal minors $L_+$ of $L_{\bullet}$ in Definition [\ref{MinorGeneralDefinitions}] to $L_{\bullet}=L_{\bullet}(v; \ebar)$ from Definition [\ref{DEFLaxOpIntro}].  

\begin{definition} \label{DEFLaxOpPrincipalMinorIntro} The {principal minor} $L_+^{\perp}(v; \ebar)$ of the Lax operator $L_{\bullet}(v; \ebar)$ is its restriction to the closed subspace $H_{+} \subset H_{\bullet}$ of periodic Hardy space spanned by $\{|h \rangle = e^{\textnormal{\textbf{i}} hx} \}_{h=1}^{\infty}$ for $h=1,2,\ldots$.
\end{definition}
\begin{definition} \label{DEFLaxOpEmbeddedPrincipalMinorIntro} The {embedded principal minor} $L_+(v; \ebar)$ of the Lax operator $L_{\bullet}(v; \ebar)$ in periodic Hardy space $H_{\bullet}$ is the operator defined in block diagonal form by 
$L_+(v; \ebar) = 0 \oplus L_+^{\perp}(v; \ebar)$ with respect to the decomposition $H_{\bullet} = H_0 \oplus H_+$, where $L_+^{\perp}(v; \ebar)$ is the principal minor of $L_{\bullet}(v; \ebar)$. \end{definition}
\noindent By Definition [\ref{DEFLaxOpEmbeddedPrincipalMinorIntro}], the restriction of $L_{+}(v;\ebar)$ to the dense subspace $\C[w] \subset H_{\bullet}$ is \begin{equation}\label{LaxEmbeddedPrincipalMinor} L_{+}(v; \ebar) \big |_{\C[w]} =  \begin{bmatrix} 
\ \ \ \ \ \ \ 0  \ \ \ \ \ \ \ & 0 & 0 & 0 & \cdots &  \\ 
0& (-1 \ebar+ V_0)  & V_{-1} & V_{-2} &  \ddots &  \\
0 & V_{1} &  (-2 \ebar+ V_0)  & V_{-1} & \ddots & \\
0& V_2 & V_1 &(-3 \ebar + V_0 )& \ddots &   \\ 
  \vdots & \ddots & \ddots & \ddots & \ddots    \\ 

  \end{bmatrix} \end{equation}  
  
  \noindent By inspection of (\ref{LaxEmbeddedPrincipalMinor}), the principal minor $L_+^{\perp} ( v; \ebar) \cong L_{\bullet}(v - \ebar; \ebar)$ is unitarily equivalent to the Lax operator whose symbol is shifted by $-\ebar$.  For further discussion of this shift relation see \textsection 5 in \cite{Moll2}.

\subsection{Classical Benjamin-Ono Lax Operator: Spectral Shift Functions and Interlacing Spectra}

\noindent To apply results from \textsection [\ref{SECSpectralMeasuresSSF}] in \textsection [\ref{DAPproof}], we first compare the spectrum of the Lax operator $L_{\bullet}(v; \ebar)$ to that of its embedded principal minor $L_+(v; \ebar)$.  Since $\ebar >0$ in (\ref{ClassicalLaxINTRINSIC}), $-L_{\bullet}(v; \ebar)$ is elliptic.  Since $v$ is bounded, $L_{\bullet}(v; \ebar)$ is bounded perturbation of $- \ebar D_{\bullet}$ which has compact resolvent which implies: \begin{lemma} \label{GeneralizedToeplitzSpectrumDiscrete} \textnormal{[Boutet de Monvel-Guillemin \cite{DeMonvelGuillemin}]} $L_{\bullet}(v;\ebar)$ has discrete spectrum in $H_{\bullet}$ \begin{equation} \label{OriginalSpectrum} \cdots \leq C_2^{\uparrow}(v; \ebar) \leq C_1^{\uparrow}(v; \ebar) \leq C_0^{\uparrow}(v; \ebar)  \end{equation} \noindent with eigenvalues $\{C_h^{\uparrow}(v; \ebar)\}_{i=0}^{\infty} $ bounded above with $-\infty$ as the only point of accumulation. \end{lemma}

\noindent By the same argument for Proposition [\ref{GeneralizedToeplitzSpectrumDiscrete}], for the embedded principal minor $L_+(v; \ebar)$ we have:
\begin{lemma} \label{GeneralizedToeplitzSpectrumDiscrete2} $L_+(v; \ebar)$ has discrete spectrum $\{C_h^{\downarrow}(v; \ebar)\}_{h=0}^{\infty}$ in $H_{\bullet}$ with $L^{\perp}(v; \ebar)$ eigenvalues \begin{equation} \label{PMSpectrum} \cdots \leq C_2^{\downarrow}(v; \ebar) \leq C_1^{\downarrow}(v; \ebar) \end{equation} \noindent and zero eigenvalue $C_0^{\downarrow}(v; \ebar) = 0$ associated to $H_0 = \C | 0 \rangle$ in $H_{\bullet} = H_0 \oplus H_+$ for $| 0 \rangle = e^{\textnormal{\textbf{i}}0 x} = 1$. \end{lemma} \noindent We now prove that the eigenvalues found in Lemma [\ref{GeneralizedToeplitzSpectrumDiscrete}] and Lemma [\ref{GeneralizedToeplitzSpectrumDiscrete2}] interlace.  \begin{proposition} \label{InterlacingCorollary} \textcolor{black}{For $\ebar>0$ and bounded $v$,} the spectra \textnormal{(\ref{OriginalSpectrum})}, \textnormal{(\ref{PMSpectrum})} of $L_{\bullet}(v; \ebar)$, $L_+^{\perp}(v; \ebar)$ interlace \begin{equation} \label{Interlacing} \cdots \leq C_2^{\uparrow}(v; \ebar) \leq C_2^{\downarrow}(v; \ebar) \leq C_1^{\uparrow}(v; \ebar) \leq C_1^{\downarrow}(v; \ebar) \leq C_0^{\uparrow}(v; \ebar).  \end{equation} 
 \end{proposition}
 \begin{itemize} \item \textcolor{black}{\textit{Proof:}} Specialize the spectral shift function of Definition [\ref{SSFDEFohyeah}] to $\xi(c | L_{\bullet}(v; \ebar), L_+(v; \ebar))$.  By Lemma [\ref{GeneralizedToeplitzSpectrumDiscrete}] and Lemma [\ref{GeneralizedToeplitzSpectrumDiscrete2}], the spectral shift function satisfies \begin{equation} \label{LaxSSFnew} \tfrac{1}{2} \xi ' (c | L_{\bullet}(v; \ebar), L_+(v; \ebar))= \sum_{i =0}^{\infty} \delta ( c - C_h^{\uparrow}(v; \ebar)) - \sum_{i=0}^{\infty} \delta(c - C_h^{\downarrow}(v; \ebar)) \end{equation} \noindent in the weak sense of the trace formula (\ref{LKtraceformula}).  By (\ref{LaxSSFnew}), to prove (\ref{Interlacing}) it is enough to show \begin{equation} \label{1LipCon} | \xi (c | L_{\bullet}(v ; \ebar), L_+(v; \ebar) ) | \leq 1 .\end{equation} \noindent By the inequality in \textsection 2 of Birman-Pushnitski \cite{BirPush}, (\ref{1LipCon}) holds \textcolor{black}{since} $L_{\bullet}(v; \ebar) - L_+(v; \ebar)$ is readily seen to have 1 positive and 1 negative eigenvalue.  Note  $C_0^{\downarrow}(v; \ebar)=0$ is in (\ref{LaxSSFnew}). $\square$ \end{itemize}

\noindent While the shift relation $L_+^{\perp} ( v; \ebar) \cong L_{\bullet}(v - \ebar; \ebar)$ implies $C_h^{\downarrow} (v; \ebar) = - \ebar + C_{h-1}^{\uparrow} (v; \ebar)$ and relates (\ref{PMSpectrum}) to (\ref{OriginalSpectrum}), our proof of Proposition [\ref{InterlacingCorollary}] does not use the shift relation.  As we discuss in \textsection 5 of \cite{Moll2}, the shift relation and the interlacing property (\ref{Interlacing}) imply the simplicity of the spectrum (\ref{OriginalSpectrum}).

\subsection{Dispersive Action Profiles for the Classical Nazarov-Sklyanin Hierarchy} \label{DAPproof} We now prove Proposition [\ref{ClassicalNSreexpression}] which asserts that $\Phi_0^{BA}(u | v; \ebar)$ in (\ref{ClassicalNSgeneratingINTRO}) has simple interlacing zeroes and poles.

\begin{itemize}
\item \textit{Proof of Proposition \textnormal{[\ref{ClassicalNSreexpression}]}.} For bounded $v$, $L_{\bullet}(v; \ebar)$ is essentially self-adjoint on the orbit of $| 0 \rangle = e^{\textbf{i} 0 x} = 1$ in Hardy space $H_{\bullet}$.  By (\ref{ClassicalNSbakerakhiezer}) and (\ref{ClassicalNSgeneratingINTRO}), the Nazarov-Sklyanin hierarchy \begin{equation} \label{ThisIsWhatItIs} \Phi_0^{BA}(u | v; \ebar) =  \langle 0 | \frac{1}{u - L_{\bullet}( v; \ebar) } | 0 \rangle \end{equation}
\noindent is the \textcolor{black}{Titchmarsh}-Weyl function (\ref{OPRLformula}) of the Lax operator $L_{\bullet}(v; \ebar)$.  By Theorem [\ref{OPRL}], \begin{equation} \label{SofarSofar} \ \ \ \ \ \ \ \ \Phi_0^{BA}(u | v; \ebar) = \frac{1}{u} \cdot \frac{ \det_{\bullet}(u - L_{+}(v; \ebar))}{ \det_{\bullet}(u - L_{\bullet} (v; \ebar))}. \end{equation} Using the spectral shift function, combine formula (\ref{SSFformula}) and (\ref{LaxSSFnew}) to get \begin{equation} \label{HereYouGoe} \Phi_0^{BA} (u | v; \ebar) = \prod_{h=1}^{\infty} \frac{ u - C_h^{\downarrow}(v; \ebar)}{u - C_{h-1}^{\uparrow}(v; \ebar)} \end{equation} after canceling the factor of $\tfrac{1}{u}$ in (\ref{SofarSofar}) using $C_0^{\downarrow}(v; \ebar) = 0$.  By the interlacing property (\ref{Interlacing}), cancellation in (\ref{HereYouGoe}) only occurs if $C_h^{\uparrow}(v; \ebar) = C_h^{\downarrow}(v; \ebar)$ and determines a unique subsequence $\{h_i\}_{i=1}^{\infty} \subset \mathbb{Z}_+$ for which $S_0^{\uparrow}(v; \ebar) = C_0^{\uparrow}(v; \ebar)$ and $S_i^{\uparrow}(v; \ebar) = C_{h_i}^{\uparrow}(v; \ebar)$ and $S_i^{\downarrow}(v; \ebar) = C_{h_i}^{\downarrow} ( v; \ebar)$ define a pair of strictly interlacing sequences as in (\ref{InterlacingINTROyay}) so (\ref{SimpleMeromorphicFormula}) holds.  With this choice, the left side of (\ref{AverageINTROyay}) is $C_0^{\uparrow}(v; \ebar)  - C_1^{\downarrow}(v; \ebar) + C_1^{\uparrow}(v; \ebar) - C_2^{\downarrow}(v; \ebar) + \cdots$.  By the trace formula (\ref{LKtraceformula}), this alternating sum is the relative trace \begin{equation} \text{Tr}_{H_{\bullet}}(L_{\bullet}(v; \ebar) - L_+(v; \ebar)) = \langle 0 | L_{\bullet}(v ; \ebar) - L_+(v; \ebar) | 0 \rangle = V_0 =\frac{1}{2\pi} \int_0^{2\pi} v(x) dx \end{equation} \noindent which for the explicit rank $2$ $L_{\bullet}(v; \ebar ) - L_+(v; \ebar)$ is the right side of (\ref{AverageINTROyay}). $\square$ \end{itemize}

\begin{proposition} \label{KMKstatementDAP} As a consequence of \textnormal{Proposition [\ref{KMKstatementII}]} and our proof of \textnormal{Proposition [\ref{ClassicalNSreexpression}]}, the dispersive action profile $f(c| v; \ebar)$ in \textnormal{Definition [\ref{DispersiveActionProfileDEF}]} is the unique profile $f \in \textbf{P}^{\vee}$ so that \begin{itemize}
\item The $T^{\uparrow}$-observable $T^{\uparrow}(u) |_f$ in \textnormal{(\ref{TUpObservableDefinition})} is the classical Nazarov-Sklyanin hierarchy \textnormal{(\ref{ClassicalNSgeneratingINTRO})}
\item The transition measure $\tau_f^{\uparrow}$ in \textnormal{(\ref{KMKgeneral})} is the {spectral measure} of $L_{\bullet}(v; \ebar)$ at $|0 \rangle$
\item The shifted Rayleigh function $\xi_f$ in \textnormal{(\ref{SRFformula})} is the \textcolor{black}{\textit{spectral shift function}} $\xi(c | L_{\bullet}(v; \ebar), L_+(v; \ebar))$. \end{itemize}
\end{proposition}



\section{Finite Gap Conditions for Dispersive Action Profiles} \label{SECfirstresultFINITEGAP}

\noindent In \textsection\ref{SECBakerAkhiezer}, we review properties of the Baker-Akhiezer functions from Dobrokhotov-Krichever \cite{DobrokhotovKrichever}.  In \textsection \ref{SECBakerAkhiezerIDENTIFICATION}, we identify them with the Baker-Akhiezer functions (\ref{ClassicalNSbakerakhiezer}) from Nazarov-Sklyanin \cite{NaSk2} in the special case $v  = v^{\vec{s}, \vec{\chi}}(x,t; \ebar)$ of the multi-phase solutions (\ref{BOmPhaseSolution}) of Satsuma-Ishimori \cite{SatsumaIshimori1979}.  With this identification, in \textsection \ref{SECMultiPhase} we prove Theorem [\ref{THEOREMMultiPhase}]: the multi-phase waves are finite gap.  In \textsection \ref{SECReflection}, we prove Proposition [\ref{THEOREMLaurentFiniteGapPARTIALCONVERSE}]: after reflection $v \mapsto -v$, multi-phase waves are no longer finite gap.

\subsection{Properties of Dobrokhotov-Krichever Multi-Phase Baker-Akhiezer Functions} \label{SECBakerAkhiezer} In \cite{DobrokhotovKrichever}, Dobrokhotov-Krichever derived the formula (\ref{BOmPhaseSolution}) for the Satsuma-Ishimori multi-phase solutions by associating to $2n+1$ real parameters $\vec{s} \in \R^{2n+1}$ (\ref{BOMultiPhaseParameters}) a singular rational spectral curve and, to $n$ further parameters $\vec{\chi} \in \R^n$, a pair of functions $\Psi_{\pm}(u, x | v^{\vec{s}, \vec{\chi}}; \ebar)$ each solving a non-stationary Schr\"{o}dinger equation in the same $x$ and $t$ as (\ref{CBOE}) whose time-dependent potentials are the pair of functions appearing in the previously known realization of (\ref{CBOE}) as a non-local Riemann-Hilbert problem.  These two $\Psi_{\pm}$ are branches of the classical multi-phase Baker-Akhiezer function $\Psi$ on their spectral curve.  We choose to express the time dependence of the non-stationary $\Psi_{\pm}(u,x | v^{\vec{s}, \vec{\chi}}; \ebar)$ implicitly through $v^{\vec{s}, \vec{\chi}} = v^{\vec{s}, \vec{\chi}}(x,t; \ebar)$ which solves (\ref{CBOE}) in the same time variable $t$.  We now collect a non-exhaustive list of properties of these $\Psi_{\pm}$ established in the proof of Theorem 1.1 in \cite{DobrokhotovKrichever}.
\begin{theorem} \label{DKSourceTheorem} \textnormal{[Dobrokhotov-Krichever \cite{DobrokhotovKrichever}]} Given $\ebar >0$, $\vec{s} \in \R^{2n+1}$ in \textnormal{(\ref{BOMultiPhaseParameters})}, and $\vec{\chi} \in \R^n$, the Satsuma-Ishimori multi-phase quasi-periodic solutions of the Benjamin-Ono equation \textnormal{(\ref{CBOE})} from \textnormal{\cite{SatsumaIshimori1979}} may be written by formula \textnormal{(\ref{BOmPhaseSolution})} as a rational function of exponential phases.  These formulae are determined by two finite gap solutions $\Psi_{\pm}(u, x| v^{\vec{s}, \vec{\chi}}; \ebar)$ of two different non-stationary Schr\"{o}dinger equations indexed by $+,-$ for $u \in \C \setminus \R$ and $x, t \in \R$ with the following properties:
\begin{itemize}
\item \textnormal{\textbf{(i)}} As functions of $x \in \R$, $\Psi_{\pm}(u, x| v^{\vec{s}, \vec{\chi}}; \ebar)$ extend to analytic functions of $z$ in 
\begin{equation} \C_{\pm} = \{ z \in \C \ : \ \pm \textnormal{Im}(z) > 0  \} \end{equation} \noindent the upper (+) and lower (-) half-planes which bound $\partial \C_{\pm} = \R$ at $z=x$.\\
\item \textnormal{\textbf{(ii)}}  As $\textnormal{Im}[z] \rightarrow \pm \infty$, $\Psi_{\pm}$ obey the asymptotic relations \begin{equation} \label{AsymptoteTheJoy} \Psi_{\pm}(u, z | v^{\vec{s}, \vec{\chi}} ; \ebar) \sim \Big ( 1 + O( e^{ - \frac{\alpha({\vec{s}}) \cdot \textnormal{Im}[z] }{\ebar}}  ) \Big ) e^{ \frac{\textbf{\textnormal{\textbf{i}}}}{\ebar} ( uz - u^2 t)}  \end{equation} \noindent where $\alpha({\vec{s}})= \min_i (s_i^{\uparrow} - s_i^{\downarrow})$ is the size of the smallest of the $n$ gaps.\\
\item \textnormal{\textbf{(iii)}} If $\vec{s} \in \R^{2n+1}$ are chosen so that the multi-phase solution is a $2\pi$-periodic function of $x$, there exist $\Phi_{\pm}(u, x | v^{\vec{s}, \vec{\chi}}; \ebar)$ which are both $2\pi$-periodic functions of $x$ so that \begin{equation}\label{PeriodicReduction} \Psi_{\pm}(u, x| v^{\vec{s}, \vec{\chi}}; \ebar) =\Phi_{\pm}(u, x | v^{\vec{s}, \vec{\chi}}; \ebar) e^{ \frac{\textbf{\textnormal{i}}}{\ebar} ( u x - u^2 t)}.\end{equation}

\item \textnormal{\textbf{(iv)}} The two functions $\Psi_{\pm}$ are related by the identity \begin{equation} ( - \ebar D+ L(v^{\vec{s}, \vec{\chi}})) \Psi_{+}(u, x| v^{\vec{s}, \vec{\chi}}; \ebar) = \frac{1}{T^{\uparrow}(u| \vec{s})} \Psi_{-}(u, x| v^{\vec{s}, \vec{\chi}}; \ebar) \end{equation} \noindent where $D = \tfrac{1}{\textnormal{\textbf{i}}} \frac{d}{dx}$, $L(v)$ is the operator of multiplication by $v$ for multi-phase $v^{\vec{s}, \vec{\chi}}$, and 
\begin{equation} \label{Candidate} T^{\uparrow}(u| \vec{s}) = \frac{(u - s_n^{\downarrow} ) \cdots (u - s_1^{\downarrow})}{(u - s_n^{\uparrow}) \cdots (u - s_1^{\uparrow}) ( u - s_0^{\uparrow}) } .\end{equation}
\end{itemize}
\end{theorem}

\subsection{Identification of Baker-Akhiezer Functions} \label{SECBakerAkhiezerIDENTIFICATION} The key ingredient we need in \textsection [\ref{SECMultiPhase}] is this:

\begin{proposition} \label{BAidentification} For any $\ebar >0$, $\vec{s} \in \R^{2n+1}$, and $\vec{\chi} \in \R^n$ for which \textnormal{(\ref{BOmPhaseSolution})} is $2\pi$-periodic in $x$, \begin{equation} \label{BAIDformula} \Phi^{BA}(u, e^{\textnormal{\textbf{i}} x} | v^{\vec{s}, \vec{\chi}} ; \ebar) = T^{\uparrow}(u | \vec{s})\Phi_+(u, x | v^{\vec{s}, \vec{\chi}};  \ebar) \end{equation} \noindent the classical Baker-Akhiezer function \textnormal{(\ref{ClassicalNSbakerakhiezer})} of $u \in \C \setminus \R$ from Nazarov-Sklyanin \textnormal{\cite{NaSk2}} specialized to the multi-phase solutions $v = v^{\vec{s}, \vec{\chi}}(x,t; \ebar)$ defined by \textnormal{(\ref{BOmPhaseSolution})} recovers the periodic part $\Phi_+$ of the Dobrokhotov-Krichever multi-phase Baker-Akhiezer function $\Psi_+$ from \textnormal{(\ref{PeriodicReduction})} after the identification \begin{equation} \label{TheIdentificationBaby} w = e^{\textnormal{\textbf{i}} x} \end{equation} and up to a factor $T^{\uparrow}(u | \vec{s})$ defined by \textnormal{(\ref{Candidate})} and which depends only on $u$ and $\vec{s}$.
\end{proposition}

\begin{itemize}
\item \textit{Proof of \textnormal{Proposition [\ref{BAidentification}]}:} By Definition [\ref{HardyDef}] of Hardy space, $|0 \rangle = e^{\textbf{i} 0 x} = 1 \in H_{\bullet}$ spans a 1-dimensional subspace $H_0$ with orthogonal decomposition $H_{\bullet} = H_0 \oplus H_+$ in which $H_+$ contains $w \C[w]$ as a dense subspace.  Under the identification (\ref{TheIdentificationBaby}), $H_{\bullet}$ consists of $2\pi$-periodic functions of $x \in \R$ that extend to analytic functions $\Phi_+(x)$ in the upper-half plane $\C_+$ (including constant functions in $H_0$), while $H_+$ is the subspace of functions in $H_{\bullet}$ satisfying $\lim_{\text{Im}[x] \rightarrow + \infty} \Phi_+(x)= 0$ (excluding constant functions in $H_0$).  By properties (i), (ii), and (iii) of Theorem [\ref{DKSourceTheorem}] from Dobrokhotov-Krichever \cite{DobrokhotovKrichever}, for any fixed $t, \ebar, u$, \begin{equation} \label{FactOne} \Phi_{\pm}(u, x | v^{\vec{s}, \vec{\chi}}; \ebar)\in H_0 \oplus H_{\pm} \end{equation} \noindent both with constant coefficient $1$, which for $|0 \rangle = 1$ is\begin{equation} \label{FactTwo}  \langle 0 |  \Phi_{\pm}(u, x | v^{\vec{s}, \vec{\chi}}; \ebar) \rangle = 1. \end{equation} \noindent With the relation $[-\ebar D , e^{ \frac{\textbf{\textnormal{i}}}{\ebar} ( u x- u^2 t)}] = -u$ for $D = \tfrac{1}{\textbf{i}} \frac{d}{dx}$, Theorem [\ref{DKSourceTheorem}] part (iv) becomes \begin{equation} \big (u - (-\ebar D + L(v^{\vec{s}, \vec{\chi}})) \big ) \Phi_{+}(u, x | v^{\vec{s}, \vec{\chi}}; \ebar)= \frac{1}{T^{\uparrow}(u| \vec{s}) } \Phi_{-}(u, x | v^{\vec{s}, \vec{\chi}}; \ebar). \end{equation} \noindent Now since this $\Phi_{+} \in H_0 \oplus H_+ = H_{\bullet}$, use the Szeg\H{o} projection $\uppi_{\bullet}$ to replace $\Phi_+$ with $\uppi_{\bullet} \Phi_+$, then take $\uppi_{\bullet}$ of both sides and use (\ref{FactOne}) and (\ref{FactTwo}) for $\Phi_-$ and $| 0 \rangle = 1$ to get\begin{equation} \label{OhNiceFormula} (u - L_{\bullet}(v^{\vec{s}, \vec{\chi}}; \ebar))\Phi_{+}(u, x | v^{\vec{s}, \vec{\chi}}; \ebar) = \frac{1}{T^{\uparrow}(u | \vec{s})} |0 \rangle \end{equation} \noindent with $L_{\bullet}(v^{\vec{s}, \vec{\chi}}; \ebar)$ the Lax operator.  Applying the resolvent of $L_{\bullet}(v^{\vec{s}, \vec{\chi}}; \ebar)$ to either side gives: \begin{equation}\Phi_{+}(u, x | v^{\vec{s}, \vec{\chi}}; \ebar) =  \frac{1}{T^{\uparrow}(u | \vec{s})} \cdot \frac{1}{u - L_{\bullet}(v^{\vec{s}, \vec{\chi}}; \ebar)} | 0 \rangle \end{equation} \noindent Since the image of $|0 \rangle =1$ under the resolvent of the Lax operator is the definition of the classical Baker-Akhiezer function (\ref{ClassicalNSbakerakhiezer}) in Nazarov-Sklyanin \cite{NaSk2}, we have proved (\ref{BAIDformula}). $\square$

\end{itemize}

\noindent While Proposition [\ref{BAidentification}] implies that the construction in Nazarov-Sklyanin \cite{NaSk2} is more general, the role of the non-stationary Schr\"{o}dinger equations in Dobrokhotov-Krichever \cite{DobrokhotovKrichever} remains to be understood in the setting of arbitrary initial data $v$ in (\ref{CBOE}).  The exponential phases on the diagonal of the matrix (\ref{BOmPhaseSolutionMatrix}) in their formula (\ref{BOmPhaseSolution}) from \cite{DobrokhotovKrichever} originate from the exponential phases \begin{equation} \label{TheRealDeal} K (u, x |t,  \ebar) = e^{ \frac{\textbf{i}}{\ebar} (u x - u^2 t) } \end{equation}\noindent in (\ref{AsymptoteTheJoy}) and (\ref{PeriodicReduction}).  Not only are (\ref{TheRealDeal}) highly oscillatory in the small dispersion limit $\ebar \rightarrow 0$, they acquire a phase upon translation $x \mapsto x + 2 \pi$ and are thus multi-valued functions of $w = e^{\textbf{i} x}$.  We expect (\ref{TheRealDeal}) should play a role in a non-local Bloch-Floquet theory for the Lax operator $L_{\bullet}(v; \ebar)$. 

\subsection{Multi-Phase Solutions are Finite Gap} \label{SECMultiPhase} We now prove Theorem [\ref{THEOREMMultiPhase}].

 \begin{itemize}
\item \textit{Proof of \textnormal{Theorem [\ref{THEOREMMultiPhase}]}}: Write (\ref{BAIDformula}) in Proposition [\ref{BAidentification}] in basis $\{w^h\}_{h=0}^{\infty}$ using (\ref{TheIdentificationBaby}) and identify coefficients of $w^0$.  By (\ref{ClassicalNSgeneratingINTRO}), the left $w^0$ coefficient in (\ref{BAIDformula}) is $\Phi_0^{BA}(u  | v^{\vec{s}, \vec{\chi}}; \ebar)$.  By (\ref{FactTwo}), the right $w^0$ coefficient in (\ref{BAIDformula}) is $T(u | \vec{s})$.  By (\ref{ThisIsWhatItIs}) and (\ref{Candidate}), we now have \begin{equation} \label{GoHome} \langle 0 | \frac{1}{u - L_{\bullet}( v^{\vec{s}, \vec{\chi}}; \ebar) } | 0 \rangle = \frac{(u - s_n^{\downarrow} ) \cdots (u - s_1^{\downarrow})}{(u - s_n^{\uparrow}) \cdots (u - s_1^{\uparrow}) ( u - s_0^{\uparrow}) } . \end{equation} By Definition [\ref{DispersiveActionProfileDEF}] and Proposition [\ref{KMKstatementDAP}], the left side of (\ref{GoHome}) is the $T^{\uparrow}$-observable of the multi-phase dispersive action profile $f( c| v^{\vec{s}, \vec{\chi}}; \ebar)$.  By Definition [\ref{MultiPhaseProfileDEF}] and formula (\ref{TakePartialFracOfThis}), the right side of (\ref{GoHome}) is the $T^{\uparrow}$-observable of the Dobrokhotov-Krichever profile $f( c | \vec{s})$.  Since profiles are determined by their $T^{\uparrow}$-observables (\ref{TUpObservableDefinition}), this proves (\ref{FiniteGapWhatWeNeed}). $\square$ \end{itemize}
\subsection{Reflected Multi-Phase Initial Data are not Finite Gap} \label{SECReflection} We now prove Proposition [\ref{THEOREMLaurentFiniteGapPARTIALCONVERSE}].

 \begin{lemma} \label{ComeOnLemma0} For the $1$-phase periodic traveling wave $v^{\vec{s}, \vec{\chi}}(x,t;\ebar)$ defined by \textnormal{(\ref{BO1PhaseForm})}, at some time $t$ its Fourier coefficients are all positive, but at no time $t$ are its Fourier coefficients all negative. \end{lemma} \noindent Lemma [\ref{ComeOnLemma0}] follows from (\ref{BO1PhaseForm}) or formula (5) in \cite{AmbroseWilkeningCOMPUTATION}.  The next two lemmas follow from the fact that for $B >0$, a probability measure $d \mu$ on $\R$ has support in $[-B,B]$ if and only if for all $p \in \mathbb{Z}_+$ \begin{equation} \int_{- \infty}^{+\infty} E^{2p} d\mu(E) < B^{2p} .\end{equation}
\begin{lemma} \label{ComeOnLemma3} For $\ebar>0$ and bounded $v$, the spectral measure $\tau^{\uparrow}(c | v; \ebar)$ of $L_{\bullet}(v; \ebar)$ at $|0 \rangle = 1$ is the transition measure of the dispersive action profile $f(c| v; \ebar)$, so by \textnormal{Corollary [\ref{KMKsupport}]} $f(c| v; \ebar)$ has finite-many gaps if and only if for all $p \in \mathbb{Z}_+$ there is some $B_{v; \ebar}>0$ independent of $p$ so that \begin{equation} \int_{-\infty}^{+\infty} c^{2p} d\tau^{\uparrow}(c | v; \ebar)\leq B_{v; \ebar}^{2p}. \end{equation}
\end{lemma}\begin{lemma} \label{ComeOnLemma2} For $v \in L^2(\mathbb{T})$, $v(x) = \sum_{k=-K_v}^{+K_v} V_k e^{-\textnormal{\textbf{i}} kx}$ for $K_v >0$ if and only if for all $p \in \mathbb{Z}_+$ \begin{equation} \sum_{k=1}^{\infty} k^{2p} |V_k|^2 \leq K_v^{2p} .\end{equation} \end{lemma}
\begin{itemize}
\item \textit{Proof of \textnormal{Proposition [\ref{THEOREMLaurentFiniteGapPARTIALCONVERSE}]}}: By contradiction, assume $- v^{\vec{s}, \vec{\chi}}$ is finite gap. To leading-order in $\ebar$, the coefficient $T_{\ell}^{\uparrow}(v; \ebar)$ of $u^{-\ell-1}$ in (\ref{ClassicalNSgeneratingINTRO}) is determined by the Sobolev norms of $v$: \begin{equation} \label{SobolevInNazarovSklyanin} T^{\uparrow}_{\ell} (v; \ebar) = (-\ebar)^{\ell-2} \sum_{h=1}^{\infty} h^{\ell -2} |V_h|^2 + O(\ebar^{\ell -3}).\end{equation} \noindent By Lemma [\ref{ComeOnLemma0}] and (\ref{BOmPhaseSolution}), without loss of generality all Fourier coefficients of the reflected multiphase initial data $-v^{\vec{s}, \vec{\chi}}$ are {negative}.  Since $-\ebar$ is also negative, the $O(\ebar^{\ell-3})$ term in (\ref{SobolevInNazarovSklyanin}) is $(-1)^{\ell-2} \cdot R_{\ell}$ for some $R_{\ell}>0$, which proves the first $\leq$ in \begin{equation} \label{YesIndeed}\ebar^{\ell -2} \sum_{h=1}^{\infty} h^{\ell-2} |V_h|^2 \leq  | T^{\uparrow}_{\ell}(v; \ebar) | \leq' B_{v; \ebar}^{\ell}.\end{equation} \noindent For the second $\leq'$ with $B_{v; \ebar}>0$, use the assumption $- v^{\vec{s}, \vec{\chi}}$ finite gap and Lemma [\ref{ComeOnLemma3}].  Formula (\ref{YesIndeed}) and Lemma [\ref{ComeOnLemma2}] imply $-v^{\vec{s}, \vec{\chi}}$ is Laurent in $e^{\textbf{i} x}$, contradicting (\ref{BOmPhaseSolution}). $\square$ \end{itemize}


\pagebreak

\section{Toeplitz Spectral Shift Functions and Convex Action Profiles} \label{SECToeplitz}

\noindent In this section we give a non-local characterization of the convex action profile $f(c|v;0)$ for bounded $v$ from Definition [\ref{DEFConvexActionProfile}] in Proposition [\ref{CHBHConstruction}] below.  To establish this characterization, we restate Szeg\H{o}'s First Theorem for Toeplitz operators using a spectral shift function implicit in Simon \cite{SimonSzego}.
\subsection{Rayleigh Measures of Convex Action Profiles are Push-Forwards of Uniform Measures} \label{SECConvexActionProfiles} \noindent For $\mathbb{T} = \{ w \in \C : |w|=1\}$ and any measurable $v : \mathbb{T} \rightarrow \R$, recall that the push-forward of the uniform measure $\rho_0$ on $\mathbb{T}$ along $v: \mathbb{T}\rightarrow \mathbb{R}$ is the probability measure $v_{*} \rho_0$ on $\mathbb{R}$ for which 
\begin{equation} \label{DefBagel} \int_{- \infty}^{+\infty} \phi(c) d (v_{*} \rho_0)(c) = \oint_{\mathbb{T}} \phi(v(w)) \frac{ dw}{ 2 \pi \textbf{i} w}. \end{equation}
\noindent for all bounded continuous $\phi$.  Choosing $v_{*}\rho_0$ as a Rayleigh measure as in \textsection [\ref{SECProfilesSRF}], (\ref{DefBagel}) implies:

\begin{proposition} \label{ConvexActionProfileSanity} For bounded $v$, the convex action profile $f(c | v;0)$ of \textnormal{Definition [\ref{DEFConvexActionProfile}]} is the convex profile as in \textnormal{Definition [\ref{ConvexProfileDEFINITION}]} whose Rayleigh measure $dF(c | v; 0)$ is the push-forward $v_* \rho_0$ of the uniform measure $\rho_0$ on $\mathbb{T}$ along $v: \mathbb{T} \rightarrow \R$, i.e. $d F(c| v; 0) = dF^{\uparrow} (c| v; 0) = d(v_{*} \rho_0)(c)$ with $dF^{\downarrow}(c| v; 0)=0$ and whose Rayleigh function $F(c|v;0)$ is given by \textnormal{(\ref{RayleighFunctionConvexActionProfile})}.\end{proposition} 
  \begin{figure}[htb]
\centering
\includegraphics[width=0.6 \textwidth]{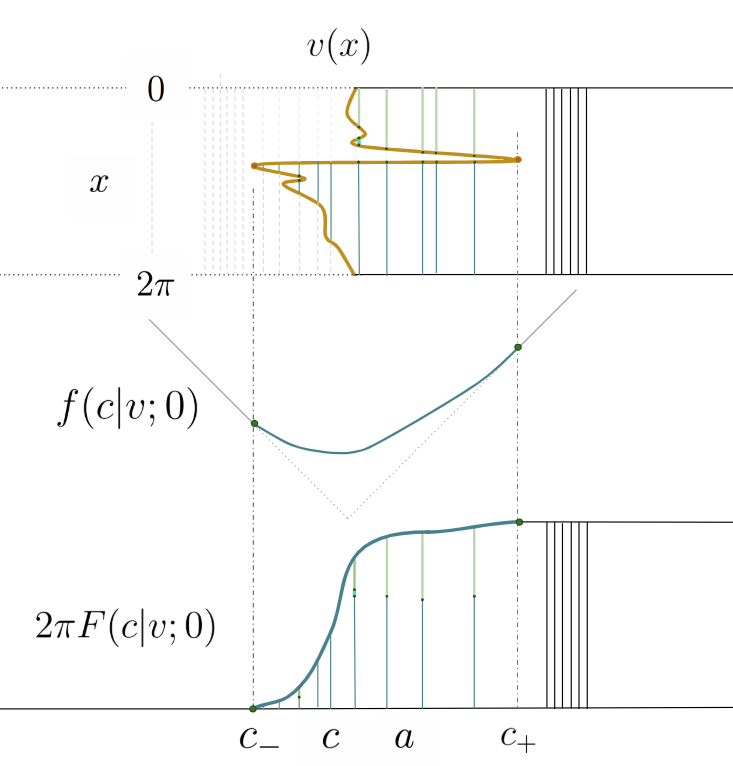}
\caption{A generic smooth $v: \mathbb{T} \rightarrow \R$ graphed horizontally and depicted with its convex action profile $f(c|v;0)$, its monotonically increasing Rayleigh function $F(c|v;0)$ with factor $2\pi$ reflecting the spatial period of $v$, and its support $[c_-, c_+]$ with $c_- = \inf_x v$ and $c_+ = \sup_x v$ containing the average $a = V_0 = \int_0^{2\pi} v(x) \tfrac{dx}{2\pi}$ as in Figure [\ref{ProfileBOSmallDispersionLimitTEMP}].  The support is connected if $v$ is continuous, a local condition on $v$.}
\end{figure}

\subsection{Multiplication Operators and Convex Action Profiles} We first give a spectral realization of the convex action profile $f(c|v;0)$ through multiplication operators.  Let $(H, \langle \cdot, \cdot \rangle)  = L^2(\mathbb{T})$.
\begin{definition} \label{MultiplicationOperatorDefinition} For bounded $2 \pi$-periodic $v$, the {multiplication operator} $L(v)$ on $H$ is defined by \begin{equation} \label{HereMult} (L(v) \Phi)(w) = v(w) \cdot \Phi(w). \end{equation}
\noindent We say that $L(v)$ is the multiplication operator with {symbol} $v$.
\end{definition} 

\begin{lemma} \label{EasyLemmaBOOPDOOP} For bounded $2\pi$-periodic real $v$, $L(v)$ is bounded and self-adjoint on $H$. 
\end{lemma}

\begin{proposition} \label{PropositionBOOP} The spectral measure of the multiplication operator $L(v)$ at $ |0 \rangle = 1 \in H$ is $dF (c|v;0) = d(v_* \rho_0)(c)$ the push-forward of the normalized uniform measure on the unit circle $\mathbb{T}$ along $v: \mathbb{T} \rightarrow \R$ and thus the Rayleigh measure of the convex action profile of \textnormal{Definition [\ref{DEFConvexActionProfile}]}.  \end{proposition} \noindent Lemma [\ref{EasyLemmaBOOPDOOP}] is standard.  Definition [\ref{SpectralMeasureDef}] and Proposition [\ref{ConvexActionProfileSanity}] imply Proposition [\ref{PropositionBOOP}].

\noindent   Proposition [\ref{PropositionBOOP}] gives a characterization of the convex action profile through the spectral theory of a local differential operator $L(v)$ of order $0$.  This perspective is relevant in the study of the small dispersion limit of the classical Korteweg-de Vries equation, since the multiplication operator $L(v)$ is the small dispersion limit of the associated classical Lax operator.  In the same manner, to streamline our description in \textsection [\ref{SECSmallDispersionLimit}] of the small dispersion asymptotics $\ebar \rightarrow 0$ of the integrable hierarchy (\ref{ClassicalNSgeneratingINTRO}) for the classical Benjamin-Ono equation (\ref{CBOE}), in \textsection [\ref{SECToeplitzSzegoFirstTheorem}] we characterize the convex action profile in terms of the small dispersion limit of the associated classical Lax operator $L_{\bullet}(v; \ebar)$.

\subsection{Toeplitz Operators and Embedded Principal Minors} \label{SECToeplitzGATHERING} We now specialize Definition [\ref{DEFLaxOpIntro}] to the $\ebar = 0$ case of Toeplitz operators \cite{BottcherSilbermann, BottcherSilbermannIntro} and discuss their spectral theory next in \textsection [\ref{SECToeplitzSzegoFirstTheorem}].  Recall from \textsection [\ref{SECnazsklyINTRO}] the Definition [\ref{HardyDef}] of the Hardy space $H_{\bullet}$ of $H = L^2(\mathbb{T})$. 
\begin{definition} \label{DEFToeplitz} For $2\pi$-periodic $v$, the {Toeplitz operator} $L_{\bullet}( v)$ on $H_{\bullet}$ with symbol $v$ is \begin{equation} L_{\bullet}(v) = \uppi_{\bullet} L(v) \uppi_{\bullet} \end{equation} \noindent where $L(v)$ is the {multiplication operator} on $H$ in \textnormal{(\ref{HereMult})} and $\uppi_{\bullet}$ is the Szeg\H{o} projection \textnormal{(\ref{HardyIntegralOperator})}. \end{definition} \noindent We write $H_{\bullet}$ for Hardy space, not $H_+$, since $H_{\bullet}$ contains $|h \rangle = e^{\textbf{i} {h} x}$ for $h=0,1,\ldots$ including $0$.  Choose $| 0 \rangle = e^{\textbf{i} 0 x} = 1 \in H_{\bullet}$ which spans $H_0 = \C| 0 \rangle$ and yields the orthogonal decomposition $H_{\bullet} = H_0 \oplus H_+$ where $H_+$ contains $w \C[w]$ as a dense subset.  Specializing the orthogonal projections $\pi_0$, $\pi_+$ from Definition [\ref{MinorGeneralDefinitions}] to the case of $L_{\bullet} = L_{\bullet}(v)$ in Hardy space $H_{\bullet}$, we have:

\begin{definition} For $2\pi$-periodic $v$, the {embedded principal minor} $L_+(v)$ of $L_{\bullet}( v)$ on $H_{\bullet}$ is \begin{equation} L_{+}(v) = \uppi_{+} L(v) \uppi_{+} \end{equation} \noindent where $L(v)$ is the {multiplication operator} $L(v)$ on $H$ and $\uppi_+$ is the shifted Szeg\H{o} projection to $H_{\bullet}$. \end{definition}  \noindent For $ V_k = \textcolor{black}{\int_0^{2 \pi}}e^{\textnormal{\textbf{i}} k x} v(x) \frac{ dx}{2\pi}$ the Fourier modes of the symbol $v(x) = \sum_{k=-\infty}^{\infty} V_{-k} e^{+ \textbf{i} k x}$, in the basis $\{e^{\textbf{i} h x}\}_{h=0}^{\infty}$ of $\C[w] \subset H_{\bullet}$, the Toeplitz operator $L_{\bullet}(v)$ and its embedded principal minor $L_+(v)$ are
\begin{equation}\label{ToeplitzMatrix} L_{\bullet}(v) \big |_{\C[w]} = \begin{bmatrix} 
V_0 & V_{-1} & V_{-2} & \cdots  \\ 
V_{1} & V_0 & V_{-1} & \ddots  \\
 V_{2} & V_{1} & V_0 & \ddots \\
  \vdots & \vdots & \ddots & \ddots \end{bmatrix} \ \ \ \ \ \ \ \ \ \ \ L_{+}(v) \big |_{\C[w]} = \begin{bmatrix} 
0 & 0 & 0 & \cdots  \\ 
0 & V_0 & V_{-1} & \ddots  \\
0 & V_{1} & V_0 & \ddots \\
  \vdots & \vdots & \ddots & \ddots \end{bmatrix}. \end{equation}
\noindent 

\subsection{Szeg\H{o}'s First Theorem and Convex Action Profiles} \label{SECToeplitzSzegoFirstTheorem}
\noindent We now give a second realization of the convex action profiles $f(c|v;0)$ in the spectral theory of Toeplitz operators.
 \begin{theorem} \label{ToeplitzSpectrumAbsolutelyContinuous} \textnormal{[Rosenblum \cite{Ros2}]} For bounded real $v$, $L_{\bullet}(v)$ has absolutely continuous spectrum.
 \end{theorem} 
 
 \noindent By this result of Rosenblum \cite{Ros2}, we can already see that the spectral theory of Toeplitz operators $L_{\bullet}(v)$ which we expect at $\ebar = 0$ is drastically different than the spectral theory of the classical Lax operator $L_{\bullet}(v; \ebar)$ for $\ebar >0$ which has discrete spectrum by Lemma [\ref{GeneralizedToeplitzSpectrumDiscrete}].  Fortunately, in \textsection [\ref{SECJacobiKerov}] we made no assumptions on the spectrum of our general self-adjoint $L_{\bullet}$.  Our only requirement was the essential self-adjointness of $L_{\bullet}$ on the orbit of a given $\psi_0$.  We now verify this property for $L_{\bullet}(v)$.

\begin{theorem} \label{Toeplitz1911} \textnormal{[Toeplitz 1911]} $L_{\bullet}(v)$ is bounded if and only if $v$ is bounded.  Moreover, the spectrum of $L_{\bullet}(v)$ coincides with the essential range of $v$ and $||L_{\bullet}(v) ||_{\textnormal{op}} = || v||_{\infty}$.
\end{theorem} \noindent For a proof of Theorem [\ref{Toeplitz1911}] see Theorem 2.7 in B\"{o}ttcher-Silbermann \cite{BottcherSilbermann}.  This result implies:
\begin{lemma} \label{ToeplitzRealNice} If $v$ is real and bounded,  $L_{\bullet}(v)  |_{\C[w]}$ is essentially self-adjoint on the orbit of $|0 \rangle = 1$.  \end{lemma}
\noindent Essential self-adjointness implies stability of the Galerkin approximations \cite{BottcherSilbermann, BottcherSilbermannIntro} and is also the key assumption in Theorem [\ref{OPRL}] which implies
\begin{equation} \label{FamousPerturbationDeterminant}\langle 0 | \frac{1}{u - L_{\bullet}(v)}  | 0 \rangle = \frac{1}{u} \frac{ \det_{H_{\bullet}} (u - L_+(v)) }{ \det_{H_{\bullet}} (u - L_{\bullet}(v)) } .\end{equation} 

\noindent The computation of the perturbation determinant in (\ref{FamousPerturbationDeterminant}) is a famous result in Toeplitz theory:

\begin{theorem} \label{SzegoFirstTheorem} \textnormal{[Szeg\H{o}'s First Theorem]} For bounded $2\pi$-periodic real $v$ the $\frac{1}{u}$ multiple of the perturbation determinant of $u$-additive shifts of the Toeplitz operator $L_{\bullet}(v)$ with respect to its embedded principal minor $L_+(v)$ is the geometric mean of $\frac{1}{u - v}$: \begin{equation} \label{SzegoFirstTheoremFormula}  \frac{1}{u} \cdot \frac{ \det_{H_{\bullet}} (u - L_{+}(v)) }{ \det_{H_{\bullet}} ( u - L_{\bullet}(v)) }  = \textnormal{exp} \Bigg ( \oint_{\mathbb{T}} \log \Bigg [ \frac{1}{u-v(w)} \Bigg ] \frac{ dw}{ 2 \pi \textnormal{\textbf{i}} w} \Bigg ). \end{equation}
\end{theorem} \noindent Szeg\H{o}'s First Theorem, also known as the ``weak Szeg\H{o} theorem," is not often stated for perturbation determinants of Toeplitz operators but instead as an asymptotic result for determinants of large Toeplitz matrices as originally conjectured by P\'{o}lya, see for example Theorem 5.10 in \textsection 5.5 of B\"{o}ttcher-Silbermann \cite{BottcherSilbermannIntro} or Theorem 2 in Deift-Its-Krasovsky \cite{DeiftItsKra}. However, as discussed by Simon in Remark 2 of Theorem 1.6.1 in \cite{SimonSzego}, in 1920 Szeg\H{o} did actually prove Theorem [\ref{SzegoFirstTheorem}] which implies the P\'{o}lya conjecture following a recommendation from Fekete.  The perturbation determinant in (\ref{SzegoFirstTheoremFormula}) coincides with the asymptotic ratio of characteristic polynomials in formula (1.6.8) of Simon \cite{SimonSzego} due to the essential self-adjointness in Lemma [\ref{ToeplitzRealNice}].\\
\\
\noindent We now realize the convex action profile $f(c|v;0)$ not by an auxiliary spectral theory of a local multiplication operator $L(v)$ as in Proposition [\ref{PropositionBOOP}] but from a non-local Toeplitz operator $L_{\bullet}(v)$:

\begin{proposition} \label{CHBHConstruction} As a consequence of \textnormal{Proposition [\ref{KMKstatementII}]} and \textnormal{Theorem [\ref{SzegoFirstTheorem}]}, the convex action profile $f(c|v;0)$ in \textnormal{Definition [\ref{DEFConvexActionProfile}]} is the unique profile $f \in \mathbf{P}^{\vee}$ so that\begin{itemize}
\item The $T^{\uparrow}$-observable $T^{\uparrow}(u)|_f $ in \textnormal{(\ref{TUpObservableDefinition})} is the geometric mean of $\tfrac{1}{u-v}$ in \textnormal{(\ref{SzegoFirstTheoremFormula})}
\item The transition measure $d \tau^{\uparrow}|_f$ in \textnormal{(\ref{KMKgeneral})} is the {spectral measure} of $L_{\bullet}(v)$ at $|0 \rangle \in H_{\bullet}$
\item The shifted Rayleigh function $\xi_f$ in \textnormal{(\ref{SRFformula})}  the \textcolor{black}{\textit{spectral shift function}} $\xi (c | L_{\bullet}(v), L_+(v))$. \end{itemize}
\end{proposition}

\section{Small Dispersion Limits of Dispersive Action Profiles} \label{SECSmallDispersionLimit}

\noindent In \textsection [\ref{SECproofTHM2}] we prove our second Theorem [\ref{THEOREMSmallDispersionLimit}], that for bounded real $v(x)$, in the small dispersion limit $\ebar \rightarrow 0$ the dispersive action profiles $f(c | v; \ebar)$ converge to the convex action profiles $f( c | v; 0)$.  In \textsection [\ref{SECwhereIreview}] we prove Proposition [\ref{PROPConvexActionProfileConserved}] that convex action profiles $f( c| v; 0)$ are invariant under (\ref{CRHE}).

\subsection{Dispersive Action Profiles at Small Dispersion are Convex Action Profiles} \label{SECproofTHM2} Using our realizations of both the dispersive and convex action profiles through spectral shift functions as key ingredients, we now prove Theorem [\ref{THEOREMSmallDispersionLimit}] as an application of Kerov's theory of profiles \cite{Ke1}.

\begin{itemize}
\item \textit{Proof of \textnormal{Theorem [\ref{THEOREMSmallDispersionLimit}]}}: As $\ebar \rightarrow 0$, the Lax operator $L_{\bullet}(v; \ebar) = - \ebar D_{\bullet} + L_{\bullet}(v)$ converges to $L_{\bullet}(v; \ebar)  \rightarrow L_{\bullet}(v)$ a Toeplitz operator $L_{\bullet}(v)$ in the strong topology.  By continuity of the von Neumann spectral theorem, resolvent matrix elements converge pointwise \begin{equation} \label{TakeALookAtMeNow} \langle 0 | \frac{1}{u - L_{\bullet}(v ; \ebar) } | 0 \rangle \rightarrow \langle 0 | \frac{1}{u - L_{\bullet}(v) } | 0 \rangle \end{equation} \noindent for $u \in \C \setminus \R$.  By Proposition [\ref{KMKstatementDAP}] and Proposition [\ref{CHBHConstruction}], formula (\ref{TakeALookAtMeNow}) is the pointwise convergence of $T^{\uparrow}$-observables of the dispersive action profile $f( c | v; \ebar)$ to the convex action profile $f( c | v; 0)$, which is exactly the desired weak convergence in Theorem [\ref{THEOREMSmallDispersionLimit}]. $\square$
\end{itemize}
\noindent Note: by results in \textsection 1.4 in Kerov \cite{Ke1}, our pointwise convergence of $T^{\uparrow}$-observables of profiles $f$ in Theorem [\ref{THEOREMSmallDispersionLimit}] implies proper weak convergence of $F^{\downarrow}, F^{\uparrow}$ in the Rayleigh functions $F_f$.  

\subsection{Convex Action Profiles and Classical Dispersionless Benjamin-Ono} \label{SECwhereIreview} \noindent By Theorem [\ref{ClassicalNStheorem}] of Nazarov-Sklyanin \cite{NaSk2} and Proposition [\ref{PROPConvexActionProfileConserved}], our second Theorem [\ref{THEOREMSmallDispersionLimit}] relates hierarchies of infinitely-many conserved quantities for the classical equations (\ref{CBOE}) and (\ref{CRHE}).  For completeness, we provide a short proof of Proposition [\ref{PROPConvexActionProfileConserved}] following the textbook of Miller \cite{MillerBOOK}.
\begin{itemize}
\item \textit{Proof of Proposition} \textnormal{[\ref{PROPConvexActionProfileConserved}]}: By Proposition [\ref{ConvexActionProfileSanity}], it is enough to check that any bounded continuous $\phi : \R \rightarrow \C$ defines a conserved quantity \begin{equation} \label{ConservationFormulaDuringProof} \frac{d}{dt} \int_0^{2 \pi}  \phi(v(x,t;0))  dx= 0 \end{equation} \noindent  for short times if $v(x,t;0)$ solves (\ref{CRHE}). The short time assumption is precisely $t < t_{v}$ where $t_v$ is the breaking time when characteristics cross discussed in \textsection 3.6.1 of Miller \cite{MillerBOOK}.  To verify (\ref{ConservationFormulaDuringProof}) it is enough to take $\phi(c) = c^{l}$ for $l=1,2,3,\ldots$ since the push-forward $v_{*} \rho_0$ of the uniform measure $\rho_0$ is bounded hence determined by its moments.  $v(x,t;0)$ is differentiable for $t< t_v$, so $ \frac{d}{dt} \int_0^{2 \pi} v(x,t;0)^l dx=0$ for all $l$ follows by direct calculation. $\square$
 \end{itemize}

 \noindent For a more illuminating and intuitive proof of Proposition [\ref{PROPConvexActionProfileConserved}], take the standard interpretation of solutions $v(x,t;0)$ to (\ref{CRHE}) as the velocity field of a continuum of infinitely-many non-interacting particles on the circle with constant uniform density $\rho(x,t;0) \equiv \rho_0$.  From this point of view, for any fixed $c$ the mass $F(c|v;0)$ of particles with velocity $\leq c$ is obviously conserved by the conservation of mass and momentum of the microscopic non-interacting particles.  This argument gives a microscopic origin for the convex action profile $f(c| v;0)$ of the macroscopic field $v$.  In the same way, as discussed in \textsection [\ref{SECmotivation}], a multi-phase solution $v= v^{\vec{s}, \vec{\chi}}(x,t; \ebar)$ is itself a system of $n$ interacting $1$-phase periodic traveling waves whose conserved asymptotic wavespeeds are the band midpoints $\tfrac{1}{2} ( s_i^{\downarrow}+ s_{i-1}^{\uparrow})$ of the dispersive action profile $f( c | v^{\vec{s}, \vec{\chi}}; \ebar) = f(c| \vec{s})$.

\pagebreak

\section{Illustration of Results for Sinusoidal Initial Data} \label{SECSinusoidal}
%

\noindent In this section we calculate the dispersive and convex action profiles for $v_{\star} (x) = 2 \cos x$ as in (\ref{SinusoidalInitialData}).

\subsection{Dispersive Action Profiles for Sinusoidal Initial Data} \label{SECSinusoidalPARTONE} We discuss the next result in \textsection [\ref{SECSinusoidalDISCUSSION}].

\begin{proposition} \label{PROPSinusoidalFunctionalEquation} For $\ebar>0$ and $v_{\star}(x) =  2 \cos x$, the dispersion action profile $f(c |v_{\star};  \ebar)$ is the profile $f$ in Nekrasov-Pestun-Shatashvilli \cite{NekPesSha} whose $T^{\uparrow}$-observable \textnormal{(\ref{TUpObservableDefinition})} is the solution of \begin{equation} \label{INTROSinusoidalDifferenceEquation}T^{\uparrow} (u + \ebar) |_f+ \frac{1}{T^{\uparrow}(u) |_f } + u = 0 \end{equation} the difference equation in \textnormal{\textsection 4} of Poghossian \textnormal{\cite{Poghossian}} which satisfies $T^{\uparrow}(u)|_f \sim u^{-1}$ as $\textnormal{Im}[u] \rightarrow + \infty$.\end{proposition}

\begin{itemize}
\item \textit{Proof of \textnormal{Proposition [\ref{PROPSinusoidalFunctionalEquation}]}}: for $v_{\star}(x) = 2 \cos x$, the Lax operator (\ref{ClassicalLaxMatrix}) is tri-diagonal
\begin{equation} \label{SinusoidalLaxTriDiagonal} L_{\bullet}(v_{\star}; \ebar) \Big |_{\C[w]}
 \begin{bmatrix} 
-0 \ebar & 1& \  & \ & \  \\ 
1& -1 \ebar  & 1 & \  &  \   \\
\  & 1 &  -2 \ebar  & 1& \  \\
\ & \ & 1 & -3 \ebar & \ddots   \\ 
  \ & \ & \ & \ddots & \ddots   \\ \end{bmatrix}  \end{equation}
\noindent with all blank entries $0$.  By definition, the dispersive action profile $f( c| v_{\star}; \ebar)$ is the unique profile whose $T^{\uparrow}$-observable is the Titchmarsh-Weyl function (\ref{OPRLformula}) of (\ref{SinusoidalLaxTriDiagonal}).  For any tri-diagonal one-sided Jacobi matrix with diagonal entries $b_0, b_1, \ldots$ and off-diagonal entries $a_0, a_1, \ldots$, the principal minor $L_+^{\perp}$ is of the same form with diagonal entries $b_h^{\perp} = b_{h+1}$ and off-diagonal entries $a_{h}^{\perp} = a_{h+1}$.  Their Titchmarsh-Weyl functions $T^{\uparrow}(u)$ and $T^{\uparrow}_{\perp}(u)$ satisfy \begin{equation} \label{GeneralJacobiMatrixIdentity} a_0^2 T^{\uparrow}_{\perp} ( u) + \frac{1}{T^{\uparrow}(u)} + u - b_0 = 0. \end{equation} \noindent For $L_{\bullet}$ in (\ref{SinusoidalLaxTriDiagonal}), $L_{+}^{\perp} \cong L_{\bullet} - \ebar \text{Id}_{H_{\bullet}}$ so (\ref{GeneralJacobiMatrixIdentity}) is equivalent to (\ref{INTROSinusoidalDifferenceEquation}).  By Theorem [\ref{KMKstatement}], the $T^{\uparrow}(u)|_f$ is also a Stieltjes transform, so $T^{\uparrow}(u)|_f$ satisfies the desired boundary condition. $\square$
\end{itemize}

\subsection{Convex Action Profiles for Sinusoidal Initial Data} \label{SECSinusoidalPARTTWO} 
\begin{proposition} \label{PROPSinusoidalSmallDispersionLimit} For $v_{\star}(x) =  2 \cos x$, the convex action profile $f(c | v_{\star}; 0)$ is  \begin{equation} \label{INTROFormulaVKLS} f(c | v_{\star};0) = \begin{cases} \tfrac{2}{\pi} (c \arcsin ( \tfrac{c}{2}) + \sqrt{4 - c^2}),  \ \ \ \ \ | c| \leq 2 \\ |c| ,    \ \ \ \ \ \ \ \  \ \ \ \ \  \ \ \ \ \  \ \ \ \ \ \ \ \ \ \  \ \ \ \ \  \ \ \ \ \  \ \ \ \ \ |c| \geq 2 \end{cases}\end{equation} the convex profile discovered by Vershik-Kerov \textnormal{\cite{KeVe}} and Logan-Shepp \textnormal{\cite{LoSh}}.
\end{proposition} 
 \begin{itemize}
\item \textit{Proof 1 of \textnormal{Proposition [\ref{PROPSinusoidalSmallDispersionLimit}]}}: Direct simplification of Definition [\ref{DEFConvexActionProfile}]. $\square$
\item \textit{Proof 2 of \textnormal{Proposition [\ref{PROPSinusoidalSmallDispersionLimit}]}}: By Theorem [\ref{THEOREMSmallDispersionLimit}] and Proposition [\ref{PROPSinusoidalFunctionalEquation}], the $\ebar \rightarrow 0$ limit of $f(c| v_{\star}; \ebar)$ is determined by the $\ebar \rightarrow 0$ limit of the functional difference equation (\ref{INTROSinusoidalDifferenceEquation}): \begin{equation} \label{WSLWSL} T^{\uparrow}(u| v_{\star};0) + \frac{1}{T^{\uparrow}(u| v_{\star};0)} + u  = 0. \end{equation} \noindent The solution of (\ref{WSLWSL}) satisfying the given boundary condition is well-known to be the Stieltjes transform of Wigner's semi-circle law.  By Example 5.2.7 in Kerov \cite{Ke1}, this law is the transition measure of the profile in Vershik-Kerov \cite{KeVe} and Logan-Shepp \cite{LoSh}. $\square$
\end{itemize}

\section{Discussion of Results and Classical Dispersive Shock Waves} \label{SECmotivation} \noindent 

\noindent \textcolor{black}{The description of solutions $v(x,t; \ebar)$ to (\ref{CBOE}) at finite time $t$ in the small dispersion limit $\ebar \rightarrow 0$ is a challenging problem in asymptotic analysis which we do not address in this paper.}  Our second Theorem [\ref{THEOREMSmallDispersionLimit}] describes the conserved quantities for (\ref{CBOE}) carried by dispersive action profiles at small dispersion, not the solutions of (\ref{CBOE}) at small dispersion.  That being said, our motivation for studying dispersive action profiles is their appearance in the Whitham approximation of solutions to (\ref{CBOE}) at small dispersion as classical dispersive shock waves which we now review.  For background, see El-Hoefer \cite{ElHoeferDSW}, Lax-Levermore-Venakides \cite{LaxLevermoreVenakidesDISPERSIVE}, Miller \cite{MillerDSW}, and Whitham \cite{WhithamBOOK}.  First, consider the midpoints of dispersive action profile bands $[S_i^{\downarrow}(v; \ebar), S_{i-1}^{\uparrow}(v; \ebar)]$: \begin{equation} \label{BandMidpoints} c_i (v; \ebar) = \tfrac{1}{2} (S_i^{\downarrow} (v; \ebar) + S_{i-1}^{\uparrow}(v; \ebar)). \end{equation}
\noindent By Theorem [\ref{THEOREMMultiPhase}], for multi-phase solutions $v  = v^{\vec{s}, \vec{\chi}}(x, t; \ebar)$, the band midpoints are $\tfrac{1}{2} (s_i^{\downarrow} + s_{i-1}^{\uparrow})$.  By the Dobrokhotov-Krichever formula (\ref{BOmPhaseSolution}), the band midpoints are manifestly the {conserved asymptotic wavespeeds} of the constituent 1-phase periodic traveling waves in $v^{\vec{s}, \vec{\chi}}(x, t; \ebar)$.  Next, as the band $[s_1^{\downarrow}, s_0^{\uparrow}]$ shrinks or merges with the other band $(- \infty, s_1^{\uparrow}]$ in (\ref{BO1PhaseForm}), \begin{equation} v^{\vec{s}, \vec{\chi}} (x,t; \ebar) \sim \begin{cases} s_1^{\uparrow} + \frac{ \ebar^2 }{ \frac{1}{2 (s_0^{\uparrow} - s_1^{\uparrow})^2} + (x - s_0^{\uparrow} t )^2} \  \ \ \ \ \ \ \ \ \ \ \ \ s_1^{\uparrow} \ < \  s_1^{\downarrow} \longrightarrow s_0^{\uparrow} \\ s_0^{\uparrow} \  \ \  \ \ \ \ \ \ \ \ \ \ \ \ \  \ \ \ \ \ \ \   \ \  \ \ \  \ \ \ \ \  \ \ \ \ \ \ \  \ \ \ \ \ s_1^{\uparrow} \longleftarrow s_1^{\downarrow} \  <  \ s_0^{\uparrow}  \\  \end{cases} \end{equation} \noindent the 1-phase solution is either a $1$-soliton or constant solution of (\ref{CBOE}).  By Theorem [\ref{THEOREMSmallDispersionLimit}], as the dispersive action profile concentrates on the convex action profile $f( c| v; 0)$, the bands are merging towards the left edge $c_- = \inf_x v$ and are shrinking towards the right edge $c_+ = \sup_x v$ of $[c_-, c_+]$.  We may now describe the Whitham approximation $v^{\textnormal{Whit}}(x,t; \ebar) \approx v(x,t; \ebar)$ of the solution of (\ref{CBOE}) at small $\ebar$ and fixed time $t$ in terms of the band midpoints (\ref{BandMidpoints}) of the dispersive action profile: at fixed time $t$ after the breaking time, $v^{\textnormal{Whit}}(x,t; \ebar)$ is a classical dispersive shock wave with
\begin{enumerate}
\item \textit{Trailing edge} of low-amplitude waves of \textcolor{black}{wavespeeds} $\approx$ $c_i(v; \ebar)$ near the \textit{edge} $c_-$
\item \textit{Oscillatory bulk} of modulated $1$-phases of \textcolor{black}{wavespeeds} $\approx$ $c_i(v; \ebar)$ in the \textit{bulk} $(c_-, c_+)$
\item \textit{Leading edge} of separated $1$-solitons of \textcolor{black}{wavespeeds} $\approx$ $c_i(v; \ebar)$ near the \textit{edge} $c_+$.
\end{enumerate}

\begin{figure}[htb]
\centering
\includegraphics[width=0.7 \textwidth]{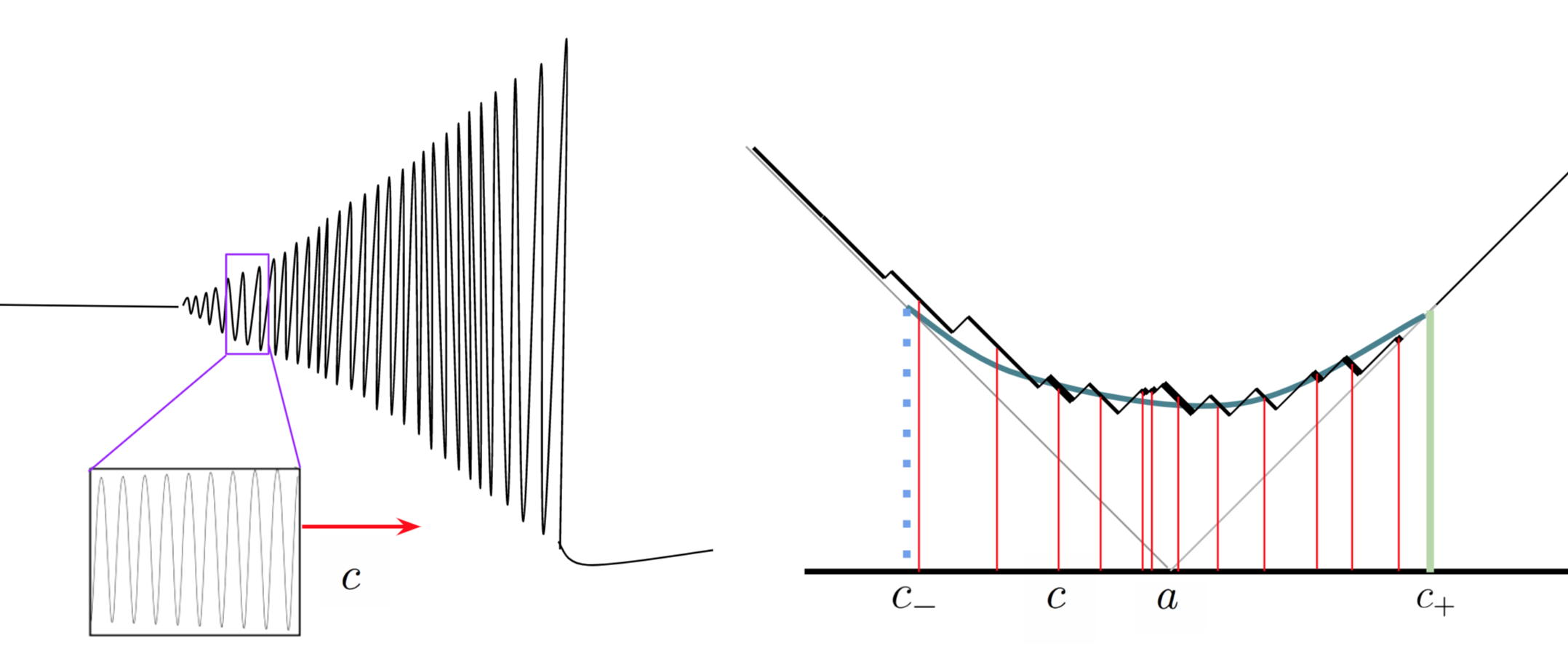}
\caption{$\square$ \textcolor{black}{Classical} dispersive shock wave $\textcolor{black}{v^{\textnormal{Whit}}(x,t; \ebar) \approx v(x,t; \ebar)}$ of modulated $1$-phase \textcolor{black}{periodic traveling wave solutions (\ref{BO1PhaseForm}) to (\ref{CBOE})} with wavespeeds $c$ together with dispersive action profile \textcolor{black}{$f( c | v; \ebar)$} \textcolor{black}{of true solution $v(x, t; \ebar)$} with band midpoints $\{c_i ( v; \ebar)\}_{i=1}^{\infty}$, \textcolor{black}{left edge $c_- = \inf_x v$, and right edge $c_+ = \sup_x v$.}}
\label{newWhithamDAPfig}
\end{figure}

\noindent \textcolor{black}{The description (1)-(3) above is adapted from the discussion in} \textsection 5.2.1 in McLaughlin-Strain \cite{McLaughlinStrain1994}.  
\noindent Profiles $f( c| v; \ebar)$ do not reflect $t$ dependence in $v^{\textnormal{Whit}} (x,t; \ebar)$ found by Dobrokhotov-Krichever \cite{DobrokhotovKrichever}.    


\noindent In simulations of Bettelheim-Abanov-Wiegmann \cite{AbBeWi2} and Miller-Xu \cite{MillerXu2011}, a localized initial data of maximum height $ \sup_x v(x,0)$ emits a highly oscillatory wave packet with maximum height $\sup_{x,t} v(x,t; \ebar) \approx 4 \sup_x v(x,0).$  We now identify frozen regions of dispersive action profiles and, conditioned on the relation (3) in the Whitham approximation to the solutions of (\ref{CBOE}) as dispersive shock waves discussed above, account for this factor of $4$ observed in simulations.
\begin{figure}[htb]
\centering
\includegraphics[width=0.9 \textwidth]{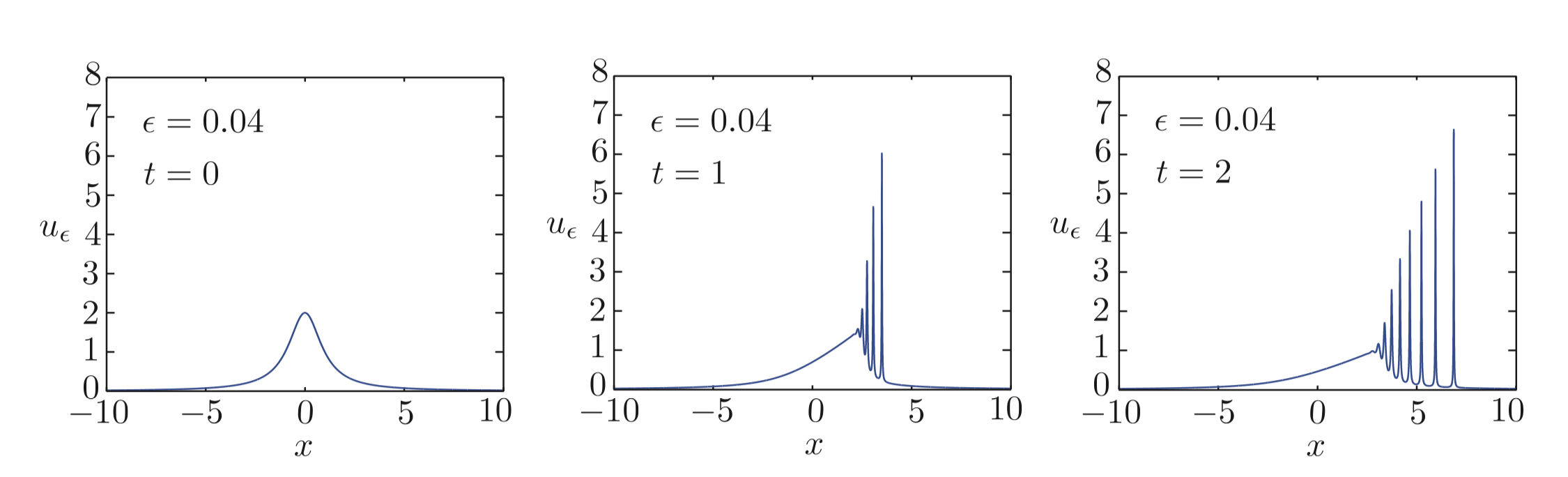}
\caption{Solution $u_{\epsilon} = v(x,t; \ebar)$ for initial data $v(x,0; \ebar ) = \tfrac{2}{1+x^2}$ at $t=0,1,2$ with $\ebar = 0.04$ from Figure 1 in Miller-Xu ``On the zero-dispersion limit of the Benjamin-Ono Cauchy problem for positive initial data" \textit{Communications in Pure and Applied Mathematics} 64(2):205-270, 2011 \cite{MillerXu2011}.}
\label{MillerXuFIGURE}
\end{figure}
\begin{proposition} \label{PROPOSITIONFrozenRegion} For $\ebar>0$ and any bounded solution $v$ of \textnormal{(\ref{CBOE})} $2\pi$-periodic in $x$ with mean $a$, the dispersive action profile $f(c| v; \ebar)$ coincides with $|c-a|$ in the region $c \in (\inf_t \sup_x v, \infty)$.\end{proposition} \begin{itemize}
\item \textit{Proof of \textnormal{Proposition [\ref{PROPOSITIONFrozenRegion}]}:} For any bounded $v$, since $||L_{\bullet}(v)||_{op} = ||v||_{\infty}$ by Toeplitz's Theorem [\ref{Toeplitz1911}] and $- \ebar D_{\bullet}$ is non-positive for $\ebar >0$, the spectrum of the classical Lax operator $L_{\bullet}(v; \ebar)$ in $H_{\bullet}$ is contained in $(- \infty, \sup_x v]$.  By the same argument, the spectrum of the embedded principal minor $L_+(v; \ebar)$ in $H_{\bullet}$ is contained in the same region, so the spectral shift function $\xi(c | L_{\bullet}(v; \ebar), L_+(v; \ebar))$ must be zero $\xi \equiv 0$ in $(\sup_x v, \infty)$.  By Theorem [\ref{ClassicalNStheorem}] and Corollary [\ref{KMKstatementDAP}], this spectral shift function is conserved for (\ref{CBOE}), so we actually have $\xi \equiv 0$ in $(\inf_t \sup_x v , \infty)$.  Since the dispersive action profile $f(c| v; \ebar)$ is determined from $\xi(c |L_{\bullet}(v; \ebar), L_+(v; \ebar))$ by Corollary [\ref{KMKstatementDAP}], the result holds. $\square$
\end{itemize}

\noindent If one views our Theorem [\ref{THEOREMSmallDispersionLimit}] as the formation of a convex \textit{limit shape} as $\ebar \rightarrow 0$, the region $(\inf_t \sup_x v, \infty)$ in Proposition [\ref{PROPOSITIONFrozenRegion}] should be called a \textit{frozen region} and $\inf_t \sup_x v$ a \textit{hard edge}.  In Figure [\ref{ProfileBOSmallDispersionLimitTEMP}], we draw a solid line at $c_+ = \sup_x v$ to represent this hard edge (no bands above $c_+$) whereas we draw a dotted line at $c_- = \inf_x v$ since $c_-$ is a \textit{soft edge} (gaps may form below $c_-$).\\
\\
\noindent According to the Whitham approximation reviewed above, the oscillatory wave packet emitted by localized initial data is a dispersive shock wave whose leading edge is comprised of separated solitons with wavespeeds $s \approx c_+$ where $c_+ = \sup_x v$ is the right edge of the support of the convex action profile $f( c| v; 0)$ in our Theorem [\ref{THEOREMSmallDispersionLimit}].  By Proposition [\ref{PROPOSITIONFrozenRegion}], no bands can form to the right of $c_+$ at any $\ebar >0$, hence soliton speeds are bounded $s \leq c_+$.  Since the soliton amplitude-speed relation for (\ref{CBOE}) is $A = 4s$ as is found e.g. El-Nguyen-Smyth \cite{ElNguyenSmyth2018}, we have $A = 4 s \leq 4 c_+$, consistent with $\sup_{x,t} v(x,t; \ebar) \approx 4 \sup_x v(x,0)$ in simulations such as Figure [\ref{MillerXuFIGURE}].

\section{Comments on Results and Comparison with Previous Results} \label{SECcomments}

\subsection{Comments on Classical Edge Universality and Integrability} The classical Benjamin-Ono equation (\ref{CBOE}) appears as a universal description of many classical fluid interfaces in two spatial dimensions (such as density, vorticity, or shear) in the asymptotic regime of long wavelength and weak non-linearity.  In particular, (\ref{CBOE}) was recently shown by Bogatskiy-Wiegmann \cite{BogatskiyWiegmann} to describe the edge dynamics of a large number of point vortices of identical circulation in the inviscid incompressible Euler equations in two dimensions.  A key step in \cite{BogatskiyWiegmann} is to derive a hydrodynamic description of the collection of point vortices and prove that the resulting vortex fluid is itself dissipationless and isotropic with non-trivial odd viscosity \cite{Avron1998}.  The odd viscous forces in the bulk give rise to a boundary layer near the edge which leads to (\ref{CBOE}).\\
\\
\noindent A striking feature of two dimensional fluids with odd viscosity such as those giving rise to (\ref{CBOE}) is the existence of an explicit family of infinitely-many conserved quantities in the bulk.  For classical ideal fluids in two dimensions, recall that vorticity is frozen into the flow, thus leading to the conservation of infinitely-many Casimirs including circulation and enstrophy.  In the presence of odd viscosity, these functions of vorticity remain conserved if one introduces an additive shift to the vorticity depending on the non-uniformity of mass density and the odd viscosity coefficient.  For a recent discussion and derivation of these Casimirs, see Abanov-Can-Ganeshan-Monteiro \cite{AbanovCanGaneshanMonteiro2019}.

\subsection{Comments on Dispersion Coefficient Notation and Sinusoidal Initial Data} \label{SECSinusoidalDISCUSSION}

\noindent Our $\ebar$ in (\ref{CBOE}) is chosen to match the standard notation $\ebar = \varepsilon_1 + \varepsilon_2$ in Nekrasov's Omega background \cite{Nek1}, a gauge theory known to be related to a quantization of (\ref{CBOE}) with $\hbar = - \varepsilon_1 \varepsilon_2$.  For a recent discussion of (\ref{CBOE}) from this point of view, see \textsection 1.1.6 in Okounkov \cite{Okounkov2018ICM}.  We study this quantization of the classical periodic Benjamin-Ono equation in \cite{Moll2} where we derive exact Bohr-Sommerfeld quantization conditions on the classical multi-phase solutions (\ref{BOmPhaseSolution}).  In \textsection [\ref{SECSinusoidal}], we saw that the convex action profile $f( c|v_{\star};0)$ for $v_{\star}$ in (\ref{SinusoidalInitialData}) is the profile in Vershik-Kerov \cite{KeVe} and Logan-Shepp \cite{LoSh}, which Nekrasov-Okounkov \cite{NekOk} proved determines the Seiberg-Witten curve and prepotential of pure $U(1)$ ${N}=2$ SUSY Yang-Mills theory on $\R^4$. In \textsection [\ref{SECSinusoidal}], we also saw that the dispersive action profile $f( c | v_{\star}; \ebar)$ is the profile which determines the $\ebar$-deformed curve and twisted superpotential of this theory in Nekrasov-Shatashvili \cite{NekShat} and Poghossian \cite{Poghossian}.  As Nekrasov-Pestun-Shatashvili \cite{NekPesSha} write, for $\ebar >0$ ``the important difference is that now ... the profile ... cannot be assumed to be a smooth function.  Instead, the profile ... shall be described by an infinite series of continuous variables,'' the interlacing local extrema of $f( c| v_{\star}; \ebar)$.



 \subsection{Comments on Dispersive Action Profiles and Classical Nazarov-Sklyanin Hierarchy} \label{SECdapDISCUSSION} \noindent The problem of constructing integrable hierarchies of conserved quantities for (\ref{CBOE}) has a long history beginning in the pioneering works \cite{BockKruskal, FokasAblowitz1983, KaupLakobaMatsuno1, KaupMatsuno, Nakamura1979}.  For definitive accounts, see the books of Ablowitz-Clarkson \cite{AblCla} and Matsuno \cite{MatsunoBook}.  Without reference to boundary conditions, (\ref{CBOE}) can be rewritten through a Lax pair of Bock-Kruskal \cite{BockKruskal} as a non-local Riemann-Hilbert problem or Hirota's bilinear formalism by which Nakamura derived a hierarchy of conserved quantities \cite{Nakamura1979}, Fokas-Ablowitz an inverse scattering transform (IST) \cite{FokasAblowitz1983}, and Kaup-Matsuno \cite{KaupLakobaMatsuno1, KaupMatsuno} a simplification of the IST for real initial data.  Since the small data work of Coifman-Wickerhauser \cite{CoifmanWickerhauser}, recent progress on the IST in the rapidly-decaying case appeared in Miller-Wetzel \cite{MillerWetzel2016rationalDIRECT} and Wu \cite{Wu1, Wu2}, while a new conservation law for (\ref{CBOE}) was found by Ifrim-Tataru \cite{IfrimTataru} without the IST.  For a recent survey of research on the classical Benjamin-Ono equation (\ref{CBOE}), see Saut \cite{Saut2018}.\\
 \\
 \noindent In the periodic case, to our knowledge the first convergent construction of an integrable hierarchy came in Nazarov-Sklyanin \cite{NaSk2}.  Their result in \cite{NaSk2} builds upon their paper \cite{NaSk1} and is stronger than Theorem [\ref{ClassicalNStheorem}]: for all $u_1, u_2 \in \C \setminus \R$, the classical Baker-Akhiezer averages Poisson commute $\{ \Phi_0^{BA} (u_1 | v; \ebar) , \Phi_0^{BA} (u_2 | v; \ebar)\}_{- {1}/{2}} = 0$ for the Gardner-Faddeev-Zakharov bracket $\{ \cdot, \cdot \}_{- {1}/{2}}$ from the $L^2$ Sobolev space for $s = -\tfrac{1}{2}$, the symplectic space for (\ref{CBOE}).  Moreover, their proof follows from quantum commutativity of a distinguished quantization of $\Phi_0^{BA} (u | v; \ebar)$ with respect to $\{ \cdot, \cdot \}_{- {1}/{2}}$.  In \cite{Moll2}, we verify that the quantum results in \cite{NaSk2} reduce to our presentation of classical results of  \cite{NaSk2} in \textsection [\ref{SECnazsklyINTRO}] in \cite{Moll2}. In recent work, low regularity conservation laws were constructed by Talbut \cite{Talbut2018} and invariant measures in several works culminating in Deng-Tzvetkov-Visciglia \cite{DengTzvetkovViscigliaPART3} and Sy \cite{Sy}.  In subsequent work, G\'{e}rard-Kappeler \cite{GerardKappeler2019} independently discovered (\ref{ClassicalNSgeneratingINTRO}) and the Baker-Akhiezer function (\ref{ClassicalNSbakerakhiezer}) of Nazarov-Sklyanin \cite{NaSk2} and gave a new proof of Theorem [\ref{ClassicalNStheorem}] using a new Lax pair for the Hamiltonian flow generated by (\ref{ClassicalNSgeneratingINTRO}) for $\{ \cdot, \cdot \}_{-1/2}$.  We verify agreement between \cite{GerardKappeler2019, NaSk2} in \textsection 5 of \cite{Moll2}.  While we do not use the Hamiltonian structure of (\ref{CBOE}) below, using work of G\'{e}rard-Kappeler \cite{GerardKappeler2019} we identify gaps in dispersive action profiles with action variables in \cite{Moll2}.  

\subsection{Comments on Finite Gap Conditions and Dobrokhotov-Krichever Spectral Curves} \label{SUBSECfinitegap} Our Theorem [\ref{THEOREMMultiPhase}] \textcolor{black}{and Proposition [\ref{THEOREMLaurentFiniteGapPARTIALCONVERSE}]} \textcolor{black}{both} agree with the subsequent classification of finite gap solutions of (\ref{CBOE}) by G\'{e}rard-Kappeler \cite{GerardKappeler2019} as we \textcolor{black}{show} in \textsection 9 of \cite{Moll2}.  Our proofs rely on properties of finite gap potentials of non-stationary Schr\"{o}dinger equations from Dobrokhotov-Krichever \cite{DobrokhotovKrichever} which we recount in Theorem [\ref{DKSourceTheorem}].  The role of these non-stationary Schr\"{o}dinger equations in the study of (\ref{CBOE}) with generic periodic initial data merits further investigation.  The use of finite gap potentials for non-stationary Schr\"{o}dinger equations in \cite{DobrokhotovKrichever} builds on the construction by Krichever \cite{Krichever1977} of complex quasi-periodic finite gap potentials for non-stationary Schr\"{o}dinger equations from generic curves, Dubrovin's conditions \cite{Dubrovin1981, Dubrovin1985} for reality and smoothness of the potentials, and the later work of Dobrokhotov-Maslov \cite{DobrokhotovMaslov} and Dubrovin-Krichever-Malanyuk-Makhankov \cite{DubrovinKricheverMalanyukMakhankov}.  For background on this theory of finite gap potentials, see surveys by Krichever \cite{Krichever1991} and Matveev \cite{MatveevSurvey}.

\subsection{Comments on Small Dispersion Asymptotics} Our second Theorem [\ref{THEOREMSmallDispersionLimit}] is a result for conserved quantities, not for solutions of (\ref{CBOE}).  For the small dispersion limit of solutions of (\ref{CBOE}), recent progress for smooth rapidly-decaying initial data appeared in Miller-Xu \cite{MillerXu2011, MillerXu2012} and Miller-Wetzel \cite{MillerWetzel2016smallDISP}.  For work on the Whitham approximation for (\ref{CBOE}), see Dobrokhotov-Krichever \cite{DobrokhotovKrichever}, Matsuno \cite{Matsuno1998, Matsuno1998Second}, Jorge-Minzoni-Smyth \cite{JorgeMinzoniSmyth1999}, and El-Nguyen-Smyth \cite{ElNguyenSmyth2018}.  For an extension of Dubrovin's universality conjectures \cite{DubrovinLECTURES} to (\ref{CBOE}), see Masoero-Raimondo-Antunes \cite{MasoeroRaimondoAntunes2015}.

\subsection{Comments on Critical Regularity and Quantum Dispersive Shock Waves} 

\noindent Coincidentally, the Sobolev regularity $s=-\tfrac{1}{2}$ of the symplectic space for (\ref{CBOE}) is also the {critical regularity} of (\ref{CBOE}).  For discussion of criticality, see Saut \cite{Saut2018} and Tao \cite{Tao0B}.  Consider the critical random initial data \begin{equation} \label{PhiPhiPhi}v(x,0) = \phi(x) + \hbar^{1/2} G(x) \end{equation} \noindent where $\phi$ is subcritical, $\hbar >0$, and $G(x)$ is the mean zero log-correlated Gaussian field on $\mathbb{T}$.  In the author's thesis \cite{Moll0}, we argue that coherent state initial data in the geometric quantization of (\ref{CBOE}) by Nazarov-Sklyanin \cite{NaSk2} provides a distinguished regularization of the Cauchy problem for (\ref{CBOE}) with initial data (\ref{PhiPhiPhi}) and show that this regularization is controlled by a model of random partitions.  As in \textsection [\ref{SECmotivation}], band midpoints of profiles of these random partitions capture the random local speeds of quantum dispersive shock waves studied by Bettelheim-Abanov-Wiegmann \cite{AbBeWi2}.\\
\\
\noindent \textbf{Acknowledgments.} The author would like to thank Percy Deift, Igor Krichever, Dana Mendelson, Peter Miller, and Petar Topalov for many helpful discussions.  This work was supported by the Andrei Zelevinsky Research Instructorship at Northeastern University and by the National Science Foundation RTG in Algebraic Geometry and Representation Theory under grant DMS-1645877.


{\footnotesize

\bibliographystyle{plain}
\bibliographystyle{amsalpha}

\bibliography{Bbib2019oct28}}

\end{document}